\pdfoutput=1
\documentclass[letterpaper,twocolumn,10pt]{article}
\usepackage{include/usenix}

\usepackage{epsfig,endnotes}
\usepackage{cite}
\usepackage[cmex10]{amsmath}
\usepackage{csvsimple}
\usepackage{color} % colors for comments
\usepackage[usenames,dvipsnames,svgnames]{xcolor}
\usepackage{microtype}
\usepackage{vwcol}
\usepackage{multirow}
\usepackage{multicol}
\usepackage{pifont}
\usepackage[T1]{fontenc}
\usepackage{subcaption}
\usepackage{graphicx}
\usepackage{listings}
\usepackage{listings}
\lstdefinestyle{embedded}{
	numbers=none,
	frame=none,
	xleftmargin=0cm,
	backgroundcolor=\color{Lavender},
	framesep=1pt,
	aboveskip=3pt,
	belowskip=3pt,
}
\lstdefinestyle{small}{
	basicstyle=\linespread{0.9}\footnotesize,	
}

\mathchardef\hyphenmathcode=\mathcode`\-

\let\origlstlisting=\lstlisting
\let\endoriglstlisting=\endlstlisting
\renewenvironment{lstlisting}
    {\mathcode`\-=\hyphenmathcode
     \everymath{}\mathsurround=0pt\origlstlisting}
    {\endoriglstlisting}

\usepackage{array}
\usepackage{mdwmath}
\usepackage{mdwtab}
\usepackage{float}
\newfloat{program}{t}{lop}
\usepackage{fixltx2e}
\usepackage{xcolor}
\usepackage{xspace}

\usepackage{url}
\usepackage{breakurl}
\expandafter\def\expandafter\UrlBreaks\expandafter{\UrlBreaks\do-\do_} 

\usepackage[comma,square,sort&compress]{include/natbib}
\usepackage[implicit=false, allbordercolors=SkyBlue]{hyperref}

\usepackage{makecell}
\usepackage{include/grumble}

\newcommand{\myparagraph}[1]{\smallskip \noindent{\bf {#1}.}}
\newcommand{\myparagraphnodot}[1]{\smallskip \noindent{\bf {#1}}}
\newcommand{\code}[1]{\mbox{\texttt{#1}}}
\newcommand{\remove}{}
\newcommand{\circled}[1]{\raisebox{.5pt}{\textcircled{\raisebox{-.9pt} {#1}}}}

\newcommand{\secref}[1]{$\S$\ref{sec:#1}}

\newcommand{\figref}[1]{Figure~\ref{fig:#1}}
\newcommand{\figreftwo}[2]{Figures~\ref{fig:#1} and~\ref{fig:#2}}

\newcommand{\tabref}[1]{Table~\ref{tab:#1}}
\newcommand{\tabreftwo}[2]{Tables~\ref{tab:#1} and~\ref{tab:#2}}

\newcommand{\mpx}{Intel MPX}
\newcommand{\mpxshort}{MPX}
\newcommand{\br}{\code{\#BR}}

\newcommand*{\affaddr}[1]{#1} % No op here. Customize it for different styles.

\usepackage[inline]{enumitem}
\setlist{noitemsep,topsep=0pt,parsep=0pt,partopsep=0pt}

\usepackage{setspace}
\usepackage[margin=5pt,font={stretch=0.9}]{caption}
\begin{document}
%
%don't want date printed
\date{}
% \title{Intel MPX Explained (Technical Report)\\{\small \href{https://intel-mpx.github.io/}{https://Intel-MPX.github.io}}}
\title{Intel MPX Explained\\ {\large\normalfont An Empirical Study of Intel MPX and Software-based Bounds Checking Approaches}\\{\small\normalfont \href{https://intel-mpx.github.io/}{https://Intel-MPX.github.io}}}

\author{%
	{\em Oleksii Oleksenko$^\dag$, Dmitrii Kuvaiskii$^\dag$}\\
	{\em Pramod Bhatotia$^*$, Pascal Felber$^\ddag$, and Christof Fetzer$^\dag$}\\
	\affaddr{\small $^\dag$TU Dresden \hspace{5mm}  $^*$The University of Edinburgh \hspace{5mm}  $^\ddag$University of Neuch\^atel}
%	\email{\{A,B,C,D,E\}@university.edu}\\
}

\maketitle

% Use the following at camera-ready time to suppress page numbers.
% Comment it out when you first submit the paper for review.
%\thispagestyle{empty}

% -*- root: main.tex -*-
%!TEX root = main.tex

\subsection*{Abstract}

Memory-safety violations are a prevalent cause of both reliability and security vulnerabilities in systems software written in unsafe languages like C/C++.
%To retrofit memory safety in legacy programs, dozens of bounds-checking software-based techniques were introduced.
Unfortunately, all the existing software-based solutions to this problem exhibit high performance overheads preventing them from wide adoption in production runs.
To address this issue, Intel recently released a new ISA extension---Memory Protection Extensions (\mpx{}), a hardware-assisted full-stack solution to protect against memory safety violations. 
% such as buffer overflows and out-of-bounds accesses with low overhead.

% Recognizing the growing need for low-overhead memory safety, Intel recently released a new ISA extension---Memory Protection Extensions (\mpx{}).
% \mpx{} is a hardware-assisted full-stack solution to protect against common memory errors such as buffer overflows and out-of-bounds accesses with low overhead.
%Thanks to new bounds-checking instructions and registers, \mpx{} strives to achieve performance acceptable in production use.

In this work, we perform an exhaustive study of the \mpx{} architecture to understand its advantages and caveats.
We base our study along three dimensions: (a) performance overheads, (b) security guarantees, and (c) usability issues.
To put our results in perspective, we compare \mpx{} with three prominent software-based  approaches: (1) trip-wire---AddressSanitizer, (2) object-based---SAFECode, and (3) pointer-based---SoftBound.

Our main conclusion is that \mpx{} is a promising technique that is not yet practical for widespread adoption.
\mpx{}'s performance overheads are still high (\textasciitilde50\% on average), and the supporting infrastructure has bugs which may cause compilation or runtime errors.
Moreover, we showcase the design limitations of \mpx{}: it cannot detect temporal errors, may have false positives and false negatives in multithreaded code, and its restrictions on memory layout require substantial code changes for some programs.

{\let\thefootnote\relax\footnotetext{This paper presents only the general discussion and aggregated data; for the complete evaluation, please see the supporting website: \href{https://intel-mpx.github.io/}{https://Intel-MPX.github.io/}.
Evaluation plots and section headings have hyperlinks to the complete experimental description and results.}}
% -*- root: main.tex -*-
%!TEX root = main.tex
\section{Introduction}
\label{sec:intro}

The majority of systems software is written in low-level languages such as C or C++.
These languages allow complete control over memory layout, which is especially important for systems development.
%The absence of garbage collector is also crucial because otherwise it could lock the execution at such a critical moment as interruption handling.
%And, of course, the languages' popularity is an essential aspect; C and C++ have been among the most popular programming languages for decades.
%However, these benefits do not come without a price.
Unfortunately, the ability to directly control memory often leads to violations of \emph{memory safety}, i.e., illegal accesses to unintended memory regions \cite{MemoryErrors2012}.

In particular, memory-safety violations emerge in the form of \emph{spatial} and \emph{temporal} errors.
Spatial errors---also called buffer overflows and out-of-bounds accesses---occur when a program reads from or writes to a different memory region than the one expected by the developer.
Temporal errors---wild and dangling pointers---appear when trying to use an object before it was created or after it was deleted.
% and even when they do, it is hard to trace back from a program failure to the source of it.

These memory-safety violations may result in sudden crashes, data losses, and other nasty bugs \cite{MemoryErrors2012}.
Moreover, these vulnerabilities can also be exploited to build a \emph{memory attack}---a scenario when an adversary gets access to an illegal region of memory and can hi-jack the system or steal confidential data.
This attack vector is prevailing among low-level languages, with almost 1,200 memory vulnerabilities published only in 2016 according to the US National Vulnerability Database \cite{NVD}.

%Memory attacks can cause the whole spectrum of consequences: starting from denial of service (DoS) and ending with full control over a victim machine.
%The famous Heartbleed bug \cite{Durumeric14}, for example, allowed attackers to read any information protected by the SSL/TLS encryption, including usernames, passwords, and cryptographic keys.

Given the importance of the problem, there are numerous solutions for enforcing memory safety in unsafe languages, ranging from static analysis to language extensions \cite{ARCHER2003,CCured2002,Valgrind2007,LBC2012,BaggyBounds2009,softbound09,CheckedC16,LowFatPointers2013,SSM2016,DieHard2006,asan12,safecode06,SGXBounds2017}.
In this work, we concentrate on \emph{deterministic dynamic bounds-checking} since it is widely regarded as the only way of defending against \emph{all} memory attacks \cite{EternalWar2013,AllYouWanted2015}.
Bounds-checking techniques augment the original unmodified program with metadata (bounds of live objects or allowed memory regions) and insert checks against this metadata before each memory access.
Whenever a bounds check fails, the program is aborted and thus the attack is prevented.
Unfortunately, state-of-the-art bounds-checking techniques exhibit high performance overhead (50--150\%) which limits their usage to development stages only.

To lower runtime overheads, Intel recently released a new ISA extension---Memory Protection Extensions (\mpx{}).
Its underlying idea is to provide hardware assistance, in the form of new instructions and registers, to software-based bounds checking, making it more efficient.

Yet, to our knowledge, there is no comprehensive evaluation of \mpx, neither from the academic community nor from Intel itself.
Therefore, the goal of this work was to analyze \mpx{} in three dimensions: performance, security, and usability.
\emph{Performance} is important because only solutions with low (up to 10--20\%) runtime overhead have a chance to be adopted in practice \cite{EternalWar2013}.
It was also crucial to investigate the root causes of the overheads to pave the way for future improvements.
\emph{Security} assessment on a set of real-world vulnerabilities was required to verify advertised security guarantees.
\emph{Usability} evaluation gave us insights on \mpx{} production quality and---more importantly---on application-specific issues that arise under \mpx{} and need to be manually fixed.
%was to perform an extensive and unbiased evaluation of \mpx.

%Moreover, the meausurements themselves would not be sufficient.
To fully explore \mpx's	 pros and cons, we put the results into perspective by comparing with existing software-based solutions.
In particular, we compared \mpx{} with three prominent techniques that showcase main classes of memory safety: trip-wire Address Sanitizer \cite{asan12}, object-based SAFECode \cite{safecode06}, and pointer-based SoftBound \cite{softbound09} (see \secref{back} for details).

Our investigation reveals that \mpx{} has high potential, but is not yet ready for widespread use.
Some of the lessons we learned are:
\begin{itemize}
 \item New \mpx{} instructions are not as fast as expected and cause up to $4\times$ slowdown in the worst case, although compiler optimizations amortize it and lead to runtime overheads of \textasciitilde50\% on average.
 \item The supporting infrastructure (compiler passes and runtime libraries) is not mature enough and has bugs, such that 3--10\% programs cannot compile/run.
 \item In contrast to other solutions, \mpx{} provides no protection against temporal errors.
 \item \mpx{} may have false positives and false negatives in multithreaded code.
 \item By default, \mpx{} imposes restrictions on allowed memory layout, such that 8--13\% programs do not run correctly without substantial code changes.
 In addition, we had to apply (non-intrusive) manual fixes to 18\% programs.
\end{itemize}

Though the first three issues can be fixed in future versions, the last two can be considered fundamental design limits.
We project that adding support for multithreading would inevitably hamper performance, and relaxing restrictions on memory layout would go against \mpx{} philosophy.

%ICC implementation has an acceptable performance overhead of $47\%$.
%However, the supporting infrastructure is still not mature: ICC version has several compiler bugs; GCC, although being more stable, has much worse overheads (roughly $150\%$); the \mpx{} runtime library does not provide wrappers for all functions of the C standard library.
%Moreover, \mpx{} currently does not support multithreading and provides no protection against temporal errors.

%Our contributions are as follows:
%\begin{itemize}
%    \item Review of design, implementation, and supporting infrastructure of \mpx{} with microbenchmarks (\secref{mpx}).
%    \item Evaluation of \mpx{} in terms of performance, security, and usability on three benchmark suites (\secref{study}).
%    \item Study of three real-world applications (\secref{casestudies}).
%    \item Discussion of issues we discovered in \mpx{} (\secref{lessons}).
%\end{itemize}

% -*- root: main.tex -*-
%!TEX root = main.tex
\section{Background}
\label{sec:back}

\begin{figure}[t]
	\centering
	\includegraphics[scale=0.52]{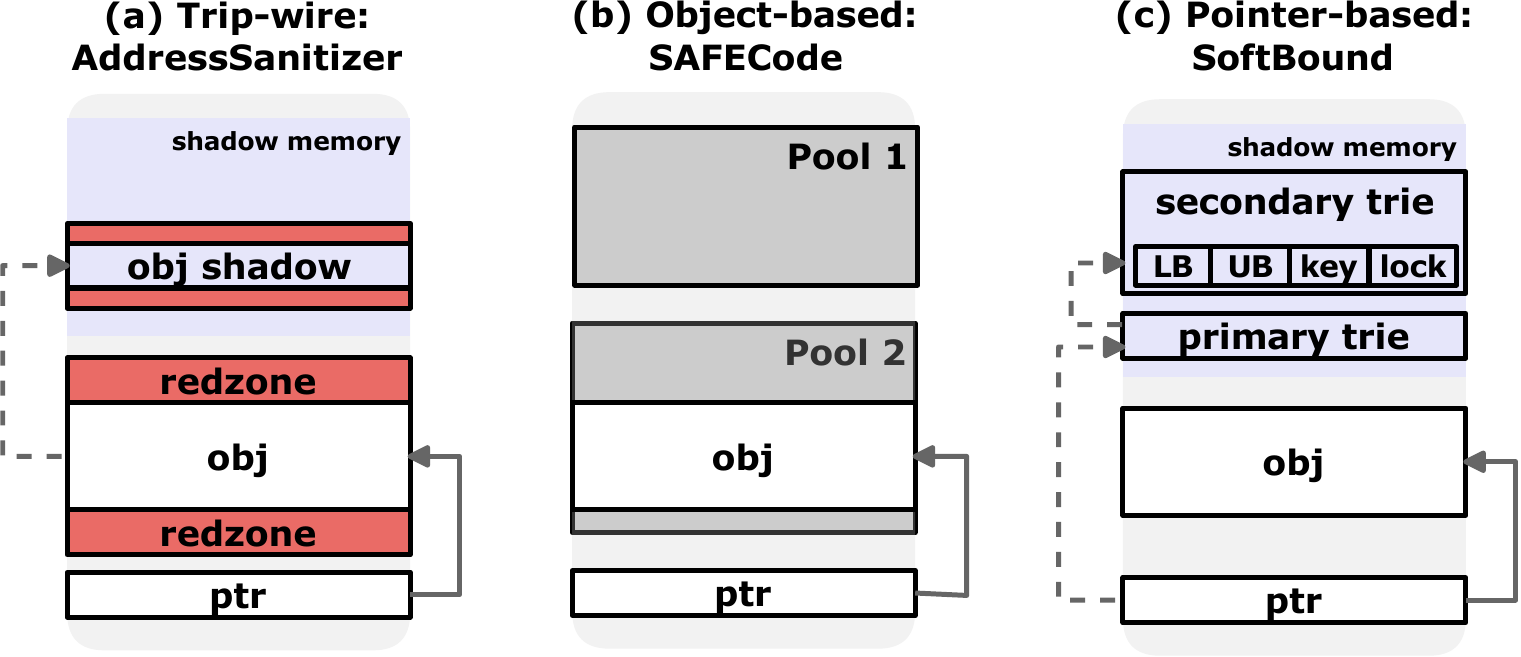}
	\caption{Designs of three memory-safety classes: trip-wire (AddressSanitizer), object-based (SAFECode), and pointer-based (SoftBound).}
	\label{fig:approaches}
\end{figure}

All spatial and temporal bugs, as well as memory attacks built on such vulnerabilities, are caused by an access to a prohibited memory region.
To prevent such bugs, \emph{memory safety} must be imposed on the program, i.e., the following invariant must be enforced: memory accesses must always stay within the originally intended (referent) objects.

Memory safety can be achieved by various methods, including pure static analysis \cite{CSSV2003,ARCHER2003}, hardware-based checking \cite{CHERI2014,LowFatPointers2013,WatchdogLite2014,MemTracker2007}, probabilistic methods \cite{DieHard2006,Exterminator2008,Archipelago2008}, and extensions of the C/C++ languages \cite{CCured2002,Cyclone2002,CheckedC16}.
In this work, we concentrate on deterministic runtime bounds-checking techniques that transparently instrument legacy programs (\mpx{} is but one of them).
These techniques provide the highest security guarantees while requiring little to no manual effort to adapt the program.
For a broader discussion, please refer to \cite{EternalWar2013}.

Existing runtime techniques can be broadly classified as trip-wire, object-based, and pointer-based \cite{AllYouWanted2015}.
In a nutshell, all three classes create, track, and check against some bounds metadata kept alongside original data of the program.
Trip-wire approaches create ``shadow memory'' metadata for the whole available program memory, pointer-based approaches create bounds metadata per each pointer, and object-based approaches create bounds metadata per each object.

For comparison with \mpx{}, we chose a prominent example from each of the aforementioned classes: AddressSanitizer, SAFECode, and SoftBound. \figref{approaches} highlights the differences between them.

\myparagraphnodot{Trip-wire approach: AddressSanitizer} \cite{asan12}.
This class surrounds all objects with regions of marked (poisoned) memory called \emph{redzones}, so that any overflow will change values in this---otherwise invariable---region and will be consequently detected.
In particular, AddressSanitizer reserves 1/8 of all virtual memory for the \emph{shadow memory} which is accessed only by the instrumentation and not the original program.
AddressSanitizer updates data in shadow memory whenever a new object is created and freed, and inserts checks on shadow memory before memory accesses to objects.
The check itself looks like this:
\begin{lstlisting}[style=embedded]
    shadowAddr = MemToShadow(ptr)
    if (ShadowIsPoisoned(shadowAddr))
            ReportError()
\end{lstlisting}
In addition, AddressSanitizer provides means to detect temporal errors via a \emph{quarantine zone}: if a memory region has been freed, it is kept in the quarantine for some time before it becomes allowed for reuse.

AddressSanitizer was built for debugging purposes and is not targeted for security (though it could be used in this context \cite{ASAP2015,Tor2017,AllYouWanted2015}).
For example, it may not detect non-contiguous out-of-bounds violations.
Nevertheless, it detects many spatial bugs and significantly raises the bar for the attacker.
It is also the most widely-used technique in its class, comparing favorably to other trip-wire techniques such as Light-weight Bounds Checking \cite{LBC2012}, Purify \cite{Purify1991}, and Valgrind \cite{Valgrind2007}.

\myparagraphnodot{Object-based approach: SAFECode} \cite{Dhurjati2006,safecode06}.
This class's main idea is enforcing the intended referent, i.e., making sure that pointer manipulations do not change the pointer's referent object.
In SAFECode, this rule is relaxed: each object is allocated in one of several fine-grained partitions---\emph{pools}---determined at compile-time using pointer analysis; the pointer must always land into the predefined pool.
This technique allows powerful optimizations and simple runtime checks against the pool bounds:
\begin{lstlisting}[style=embedded]
    poolAddr = MaskLowBits(ptr)
    if (poolAddr not in predefinedPoolAddrs)
            ReportError()
\end{lstlisting}
On the downside, SAFECode provides worse guarantees than AddressSanitizer---buffer overflow to an object in the same pool will go undetected.

We also inspected and discarded other object-based approaches.
CRED \cite{Ruwase2004} has huge performance overheads, mudflap \cite{mudflap2016} is deprecated in newer versions of GCC, and Baggy Bounds Checking \cite{BaggyBounds2009} is not open sourced.

\myparagraphnodot{Pointer-based approach: SoftBound} \cite{softbound09,cets2010}.
Such approaches keep track of pointer bounds (the lowest and the highest address the pointer is allowed to access) and check each memory write and read against them.
Note how SoftBound associates metadata \emph{not} with an object but rather with a pointer to the object.
This allows pointer-based techniques to detect intra-object overflows (one field overflowing into another field of the same struct) by \emph{narrowing bounds} associated with the particular pointer.

\mpx{} closely resembles SoftBound; indeed, a hardware-assisted enhancement of SoftBound called WatchdogLite shares many similarities with \mpx{} \cite{WatchdogLite2014}. \remove{}
For our comparison, we used the SoftBound+CETS version which keeps pointer metadata in a two-level trie---similar to \mpxshort{}'s bounds tables---and introduces a scheme to protect against temporal errors \cite{cets2010}.
The checks in this version are as follows:
\begin{lstlisting}[style=embedded]
    LoBound,UpBound,key,lock = TrieLookup(ptr)
    if (ptr < LoBound or ptr > UpBound or key != *lock)
            ReportError()
\end{lstlisting}

As for other pointer-based approaches, MemSafe \cite{Memsafe2013} is not open sourced, and CCured \cite{CCured2002} and Cyclone \cite{Cyclone2002} require manual changes in programs.

% -*- root: main.tex -*-
%!TEX root = main.tex
\section{\href{https://intel-mpx.github.io/}{Intel Memory Protection Extensions}}
\label{sec:mpx}

Intel Memory Protection Extensions (\mpx{}) was first announced in 2013 \cite{mpxintro2013} and introduced as part of the Skylake microarchitecture in late 2015 \cite{skylake2015}.
The sole purpose of \mpx{} is to transparently add bounds checking to legacy C/C++ programs.
Consider a code snippet in \figref{mpx_example}a.
The original program allocates an array \code{a[10]} with 10 pointers to some buffer objects of type \code{obj} (Line 1).
Next, it iterates through the first \code{M} items of the array to calculate the sum of objects' length values (Lines 3--8).
In C, this loop would look like this:
\begin{lstlisting}[style=embedded]
    for (i=0; i<M; i++)    total += a[i]->len;
\end{lstlisting}
Since M is a variable, a bug or a malicious activity may set M to a value that is larger than \code{obj} size and an overflow will happen.
Also, note how the array item access \code{a[i]} decays into a pointer \code{ai} on Line 4, and how the subfield access decays to \code{lenptr} on Line 6.

% -*- root: ../main.tex -*-
%!TEX root = ../main.tex
\begin{figure}[t]
\vspace{-1mm}
\begin{minipage}[t]{0.95\columnwidth}
\begin{lstlisting}[name=mpx_example1,frame=t,numbers=none,title={(a) Original code},label={lst:mpx_example1}]
struct obj { char buf[100];  int len }
\end{lstlisting}
\begin{lstlisting}[name=mpx_example2,frame=b]
obj* a[10]                      |\hfill\color{Gray}{\itshape ~;; Array of pointers to objs}|  
total = 0
for (i=0; i<M; i++):            |\hfill\color{Gray}{\itshape ~;; M may be greater than 10}|
    ai = a + i                  |\hfill\color{Gray}{\itshape ~;; Pointer arithmetic on a}|
    objptr = load ai            |\hfill\color{Gray}{\itshape ~;; Pointer to obj at a[i]}|
    lenptr = objptr + 100       |\hfill\color{Gray}{\itshape ~;; Pointer to obj.len}|
    len = load lenptr
    total += len                |\hfill\color{Gray}{\itshape ~;; Total length of all objs}|
\end{lstlisting}
\end{minipage}
\begin{minipage}[t]{0.95\columnwidth}
\begin{lstlisting}[name=mpx_example3,title={(b) Intel MPX},frame=t,label={lst:mpx_example2}]
obj* a[10]
total = 0
\end{lstlisting}
\begin{lstlisting}[name=mpx_example3,frame=none	,backgroundcolor=\color{Lavender}]
a_b = bndmk a, a+79          |\hfill\color{Gray}{\itshape ~;; Make bounds [a, a+79]}|
\end{lstlisting}
\begin{lstlisting}[name=mpx_example3,frame=none]
for (i=0; i<M; i++):
    ai = a + i
\end{lstlisting}
\begin{lstlisting}[name=mpx_example3,frame=none,backgroundcolor=\color{Lavender}]
    bndcl a_b, ai            |\hfill\color{Gray}{\itshape ~;; Lower-bound check of a[i]}|
    bndcu a_b, ai+7          |\hfill\color{Gray}{\itshape ~;; Upper-bound check of a[i]}|
\end{lstlisting} 
\begin{lstlisting}[name=mpx_example3,frame=none]
    objptr = load ai
\end{lstlisting}
\begin{lstlisting}[name=mpx_example3,frame=none,backgroundcolor=\color{Lavender}]
    objptr_b = bndldx ai     |\hfill\color{Gray}{\itshape ~;; Bounds for pointer at a[i]}|
\end{lstlisting}
\begin{lstlisting}[name=mpx_example3,frame=none]
    lenptr = objptr + 100
\end{lstlisting}
\begin{lstlisting}[name=mpx_example3,frame=none,backgroundcolor=\color{Lavender}]
    bndcl objptr_b, lenptr   |\hfill\color{Gray}{\itshape ~;; Checks of obj.len $\rceil$}|
    bndcu objptr_b, lenptr+3 |\hfill\color{Gray}$\rfloor$|
\end{lstlisting}
\begin{lstlisting}[name=mpx_example3,frame=b]
    len = load lenptr
    total += len
\end{lstlisting}
\end{minipage}
\vspace{0mm}
\caption{Example of bounds checking using Intel MPX.}
\label{fig:mpx_example}
\vspace{0mm}
\end{figure}

\figref{mpx_example}b shows the resulting code with \mpx{} protection applied.
First, the bounds for the array \code{a[10]} are created on Line 3 (the array contains 10 pointers each 8 bytes wide, hence the upper-bound offset of 79).
Then in the loop, before the array item access on Line 8, two \mpxshort{} bounds checks are inserted to detect if \code{a[i]} overflows (Lines 6--7).
Note that since the protected load reads an 8-byte pointer from memory, it is important to check \code{ai+7} against the upper bound (Line 7).

Now that the pointer to the object is loaded in \code{objptr}, the program wants to load the \code{obj.len} subfield.
By design, \mpx{} must protect this second load by checking the bounds of the \code{objptr} pointer.
Where does it get these bounds from?
In \mpx{}, every pointer stored in memory has its associated bounds also stored in a special memory region accessed via \code{bndstx} and \code{bndldx} \mpxshort{} instructions (see next subsection for details).
Thus, when the \code{objptr} pointer is retrieved from memory address \code{ai}, its corresponding bounds are retrieved using \code{bndldx} from the same address (Line 9).
Finally, the two bounds checks are inserted before the load of the length value on Lines 11--12\footnote{Note that narrowing of bounds is not shown for simplicity, see \secref{compiler}.}.

\mpx{} requires modifications at each level of the hardware-software stack\footnote{Henceforth, we focus on 64-bit Linux-based support of \mpx{}.}:
\begin{itemize}
	\item At the \emph{hardware level}, new instructions as well as a set of 128-bit registers are added. Also, a bounds violation exception (\br{}) thrown by these new instructions is introduced.
	\item At the \emph{OS level}, a new \br{} exception handler is added that has two main functions: (1) allocating storage for bounds on-demand and (2) sending a signal to the program whenever a bounds violation is detected.
	\item At the \emph{compiler level}, new \mpx{} transformation passes are added to insert \mpxshort{} instructions to create, propagate, store, and check bounds. Additional \emph{runtime libraries} provide initialization/finalization routines, statistics and debug info, and wrappers for functions from C standard library.
	\item At the \emph{application level}, the \mpxshort{}-protected program may require manual changes due to unconventional C coding patterns, multithreading issues, or potential problems with other ISA extensions. (In some cases, it is inadvisable to use \mpx{} at all.)
\end{itemize}

In the following, we detail how \mpx{} support is implemented at each level of the hardware-software stack.

\subsection{\href{https://intel-mpx.github.io/design/\#hardware}{Hardware}}
\label{sec:hardware}

At its core, \mpx{} provides 7 new instructions and a set of 128-bit bounds registers.
The current Intel Skylake architecture provides four registers named \code{bnd0--bnd3}.
Each of them stores a lower 64-bit bound in bits 0--63 and an upper 64-bit bound in bits 64--127.

\myparagraph{\href{https://intel-mpx.github.io/design/\#hardware}{Instruction set}}
The new \mpxshort{} instructions are: \code{bndmk} to create new bounds, \code{bndcl} and \code{bndcu}/\code{bndcn} to compare the pointer value against the lower and upper bounds in \code{bnd} respectively, \code{bndmov} to move bounds from one \code{bnd} register to another and to spill them to stack, and \code{bndldx} and \code{bndstx} to load and store pointer bounds in special Bounds Tables respectively.
Note that \code{bndcu} has a one's complement version \code{bndcn} which has exactly the same characteristics, thus we mention only \code{bndcu} in the following.
The example in \figref{mpx_example}b shows how most of these instructions are used.
The instruction not shown is \code{bndmov} which serves mainly for internal rearrangements in registers and on stack.

\mpx{} additionally changes the x86-64 calling convention.
In a nutshell, the bounds for corresponding pointer arguments are put in registers \code{bnd0--bnd3} before a function call and the bounds for the pointer return value are put in \code{bnd0} before return from the function.

It is interesting to compare the benefits of hardware implementation of bounds-checking against the software-only counterpart---SoftBound in our case \cite{softbound09,cets2010}.
First, \mpx{} introduces separate bounds registers to lower register pressure on the general-purpose register (GPR) file, something that software-only approaches suffer from.
Second, software-based approaches cannot modify the calling convention and resort to function cloning, when a set of function arguments is extended to include pointer bounds.
This leads to more cumbersome caller/callee code and problems with interoperability with legacy uninstrumented libraries.
Finally, dedicated \code{bndcl} and \code{bndcu} instructions substitute the software-based ``compare and branch'' instruction sequence, saving one cycle and exerting no pressure on branch predictor.

The prominent feature of \mpx{} is its backwards-compatibility and interoperability with legacy code.
On the one hand, \mpxshort-instrumented code can run on legacy hardware because \mpx{} instructions are interpreted as NOPs on older architectures.
This is done to ease the distribution of binaries---the same \mpxshort-enabled program/library can be distributed to all clients.
On the other hand, \mpx{} has a comprehensive support to interoperate with unmodified legacy code: (1) a \code{BNDPRESERVE} configuration bit allows to pass pointers without bounds information created by legacy code, and (2) when legacy code changes a pointer in memory, the later \code{bndldx} of this pointer notices the change and assigns always-true (\code{INIT}) bounds to it.
In both cases, the pointer created/altered in legacy code is considered ``boundless'': this allows for interoperability but also creates holes in \mpx{} defense\footnote{\emph{x264} from PARSEC highlights this issue: its \code{x264\_malloc} function internally calls \code{memalign} which has no corresponding wrapper. Thus, the pointer returned by this function is ``boundless''. Since all dynamic objects are created through this function, the whole program operates on ``boundless'' pointers, rendering \mpx{} protection utterly useless.} \cite{mpxFalsePositivesWithUninstrumentedCode}.

\myparagraph{\href{https://intel-mpx.github.io/design/\#boundstore}{Storing bounds in memory}}
The current version of \mpx{} has only 4 bounds registers, which is clearly not enough for real-world programs---we will run out of registers even if we have only 5 distinct pointers.
Accordingly, all additional bounds have to be stored (spilled) in memory, similar to spilling data out of general-purpose registers.
A simple and relatively fast option is to copy them directly into a compiler-defined memory location (on stack) with \code{bndmov}.
However, it works only inside a single stack frame: if a pointer is later reused in another function, its bounds will be lost.
To solve this issue, two instructions were introduced---\code{bndstx} and \code{bndldx}.
They store/load bounds to/from a memory location derived from the address of the pointer itself (see \figref{mpx_example}b, Line 9), thus making it easy to find pointer bounds without any additional information, though at a price of higher complexity.

\begin{figure}[t]
    \centering
    \includegraphics[scale=0.4]{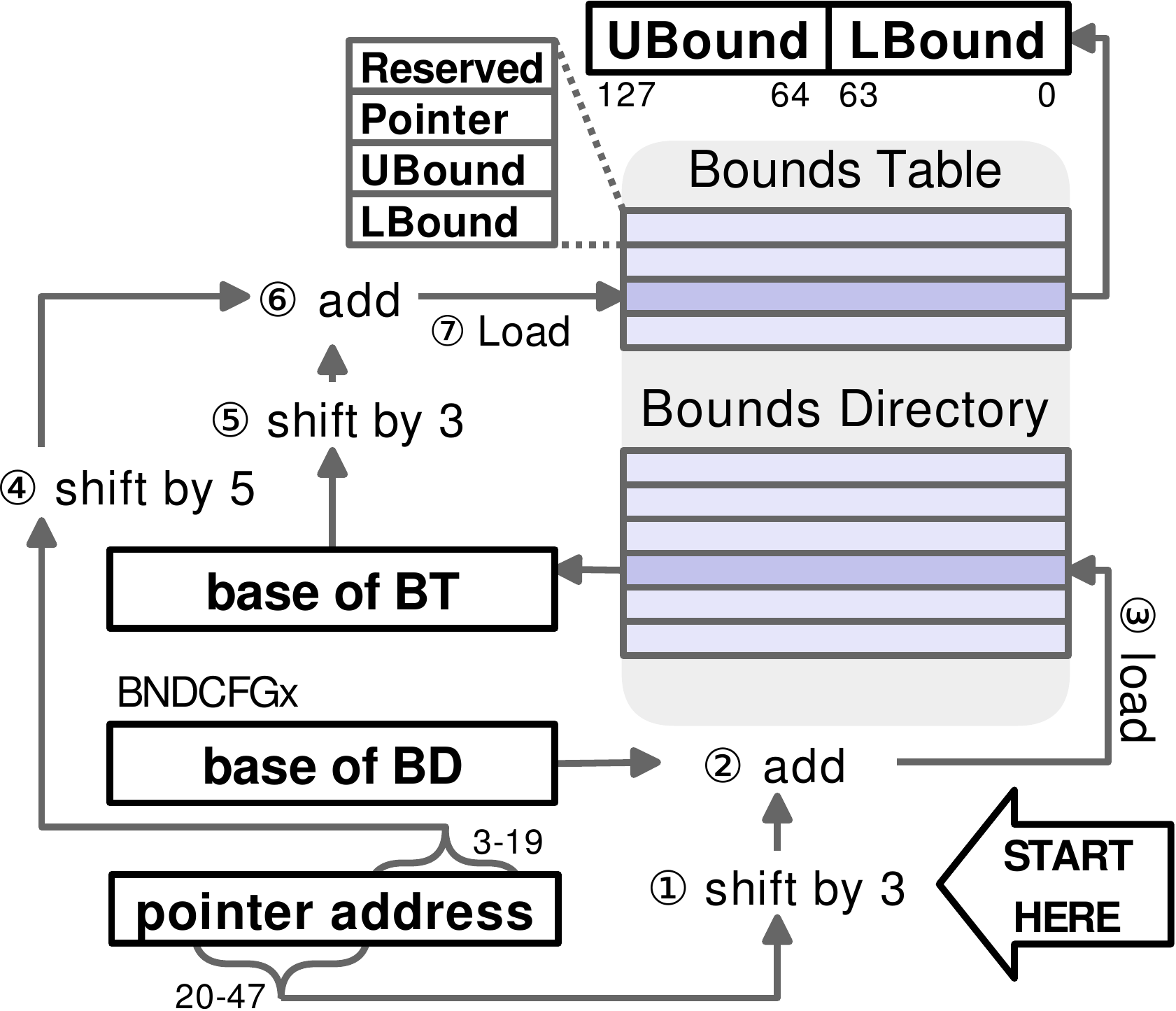}
    \caption{Loading of pointer bounds using two-level address translation.}
    \label{fig:bndld}
\end{figure}

When \code{bndstx} and \code{bndldx} are used, bounds are stored in a memory location calculated with two-level address translation scheme, similar to virtual address translation (paging).
In particular, each pointer has an entry in a Bounds Table (BT), which is allocated dynamically and is comparable to a page table.
Addresses of BTs are stored in a Bounds Directory (BD), which corresponds to a page directory in our analogy.
For a specific pointer, its entries in the BD and the BT are derived from the memory address in which the pointer is stored.

Note that our comparison to paging is only conceptual; the implementation side differs significantly.
Firstly, the MMU is not involved in the translation and all operations are performed by the CPU itself.
Secondly and most importantly, \mpx{} does not have a dedicated cache (such as a TLB cache), thus it has to share normal caches with application data.
In some cases, it may lead to severe performance degradation caused by cache thrashing.

The address translation is a multistage process.
Consider loading of pointer bounds (\figref{bndld}).
In the first stage, the corresponding BD entry has to be loaded.
For that, the CPU: \circled{1} extracts the offset of BD entry from bits 20--47 of the pointer address and shifts it by 3 bits (since all BD entries are $2^3$ bits long), \circled{2} loads the base address of BD from the \code{BNDCFGx}\footnote{In particular, \code{BNDCFGU} in user space and \code{BNDCFGS} in kernel mode.} register, and \circled{3} sums the base and the offset and loads the BD entry from the resulting address.

In the second stage, the CPU: \circled{4} extracts the offset of BT entry from bits 3--19 of the pointer address and shifts it by 5 bits (since all BT entries are $2^5$ bits long), \circled{5} shifts the loaded entry---which corresponds to the base of BT---by 3 to remove the metadata contained in the first 3 bits, and    \circled{6} sums the base and the offset and \circled{7} finally loads the BT entry from the resulting address.
Note that a BT entry has an additional ``pointer'' field---if the actual pointer value and the value in this field mismatch, \mpx{} will mark the bounds as always-true (\code{INIT}).
This is required for interoperability with legacy code and only happens when this code modifies the pointer.

This operation is expensive---it requires approximately 3 register-to-register moves, 3 shifts, and 2 memory loads.
On top of it, since these 2 loads are non-contiguous, the protected application has worse cache locality.

% Why is this figure here? It must be on the same page as fig:bndld
\begin{figure}[t]
    \centering
    \includegraphics[scale=1.2]{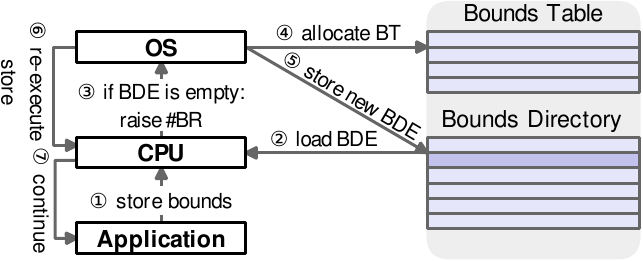}
    \caption{The procedure of Bounds Table creation.}
    \label{fig:bt-alloc}
\end{figure}

\myparagraph{\href{https://intel-mpx.github.io/microbenchmarks/\#isa}{Interaction with other ISA extensions}}
\mpx{} can cause issues when used together with other ISA extensions, e.g., Intel TSX and Intel SGX.
\mpx{} may cause transactional aborts in some corner cases when used inside an Intel TSX hardware transaction (see \cite{intelsys} for the details).
Also, since Bounds Tables and \br{} exceptions are managed by the OS, \mpx{} cannot be used as-is in an Intel SGX enclave environment.
Indeed, the malicious OS could tamper with these structures and subvert correct \mpxshort{} execution.
To prevent such scenarios, \mpx{} allows to move this functionality into the SGX enclave and verify every OS action \cite{SGXBounds2017}.
Finally, we are not aware of any side-channel attacks that could utilize \mpx{} inside the enclave.

\myparagraph{\href{https://intel-mpx.github.io/microbenchmarks/\#mpxinstr}{Microbenchmark}}
As a first step in our evaluation, we analyzed latency and throughput of \mpxshort{} instructions.
For this, we extended the scripts used to build Agner Fog's instruction tables---a de-facto standard for evaluating CPU instructions \cite{Fog11}.
For each run, we initialize all \code{bnd} registers with dummy values to avoid interrupts caused by failed bound checks.

% -*- root: ../main.tex -*-
\begin{table}[t]
\footnotesize
\centering
\begin{tabular}{p{1.49cm} p{4.3cm} p{0.4cm} p{0.34cm}}
\multicolumn{1}{c}{\bfseries Instruction} & \multicolumn{1}{c}{\bfseries Description} & \multicolumn{1}{c}{\bfseries Lat} & \multicolumn{1}{c}{\bfseries Tput} \\
\hline
\hline
\code{bndmk b,m}  & create pointer bounds              & 1 & 2 \\
\\ [-8pt]
\code{bndcl b,m}  & check mem-operand addr against lower & 1 & 1 \\
\code{bndcl b,r}  & check reg-operand addr against lower & 1 & 2 \\
\code{bndcu b,m}  & check mem-operand addr against upper & 1 & 1 \\
\code{bndcu b,r}  & check reg-operand addr against upper & 1 & 2 \\
\\ [-8pt]
\code{bndmov b,m} & move pointer bounds from mem        & 1 & 1 \\
\code{bndmov b,b} & move pointer bounds to other reg    & 1 & 2 \\
\code{bndmov m,b} & move pointer bounds to mem          & 2 & 0.5 \\
\\ [-16pt]
\code{bndldx b,m} & load pointer bounds from BT         & 4-6 & 0.4 \\
\\ [-18pt]
\code{bndstx m,b} & store pointer bounds in BT          & 4-6 & 0.3 \\
\hline
\\ [-8pt]
\multicolumn{4}{r}{\scriptsize{Note: \code{bndcu} has a one's complement version \code{bndcn}, we skip it for clarity}}
\\
\end{tabular}
\caption{Latency (cycles/instr) and Tput (instr/cycle) of \mpx{} instructions; \code{b}---\mpxshort{} bounds register; \code{m}---memory operand; \code{r}---general-purpose register operand.}
\label{tab:asm_results}
%\vspace{2mm}
\end{table}

\begin{figure}[t]
    \centering
    \includegraphics[scale=0.45]{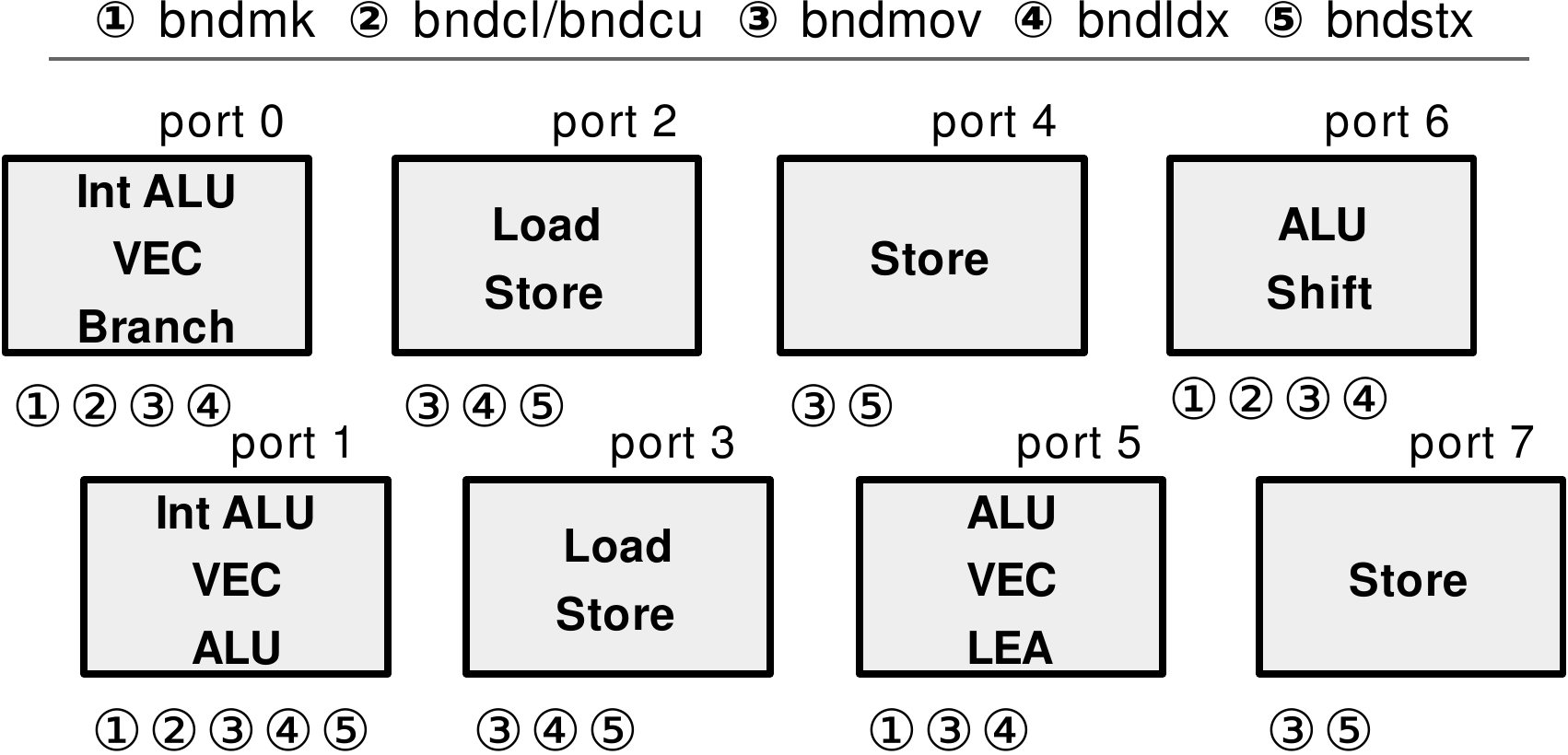}
    \caption{Distribution of \mpx{} instructions among execution ports (Intel Skylake).}
    \label{fig:microarchitecture}
\end{figure}

\tabref{asm_results} shows the latency-throughput results, and \figref{microarchitecture} depicts which execution ports can \mpxshort{} instructions use.
As expected, most operations have latencies of one cycle, e.g.,  the most frequently used \code{bndcl} and \code{bndcu} instructions.
The serious bottleneck is storing/loading the bounds with \code{bndstx} and \code{bndldx} since they undergo a complex algorithm of accessing bounds tables, as explained in the previous section.

In our experiments, we observed that \mpx{} protection does not increase the IPC (instructions/cycle) of programs, which is usually the case for memory-safety techniques (see \figref{ipc}).
This was surprising: we expected that \mpx{} would take advantage of underutilized CPU resources for programs with low original IPC.
To understand what causes this bottleneck, we measured the throughput of typical \mpxshort{} check sequences.\footnote{We originally blamed an unjustified data dependency between \code{bndcl/u} and the protected memory access (which proved wrong).}

\begin{figure*}[t]
    \centering
    \href{https://intel-mpx.github.io/microbenchmarks/#mpxchecks}{\includegraphics[scale=1.2]{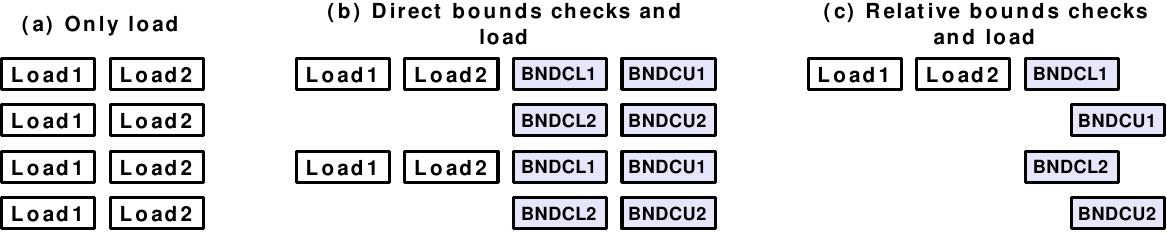}}
    \caption{Bottleneck of bounds checking illustrated: since relative memory addresses can be calculated only by port 1, a contention appears and bounds checks are executed sequentially.}
    \label{fig:mpx_checks}
\end{figure*}

\href{https://intel-mpx.github.io/microbenchmarks/\#mpxchecks}{Our measurements} pointed to a bottleneck of \mbox{``bndcl/u b,m''} instructions due to contention on port 1.
Without checks (\figref{mpx_checks} a), our original benchmark could execute two loads in parallel, achieving a throughput of 2 IPC (note that the loaded data is always in a Memory Ordering Buffer).
After adding \mbox{``bndcl/u b,r''} checks (\figref{mpx_checks} b), IPC increased to three instructions per cycle (3 IPC): one load, one lower-, and one upper-bound check per cycle.
For \mbox{``bndcl/u b,m''} checks (\figref{mpx_checks} c), however, IPC became \emph{less} than original: two loads and four checks were scheduled in four cycles, thus IPC of 1.5.
%The similar analysis applies for stores, however, the original IPC in this case is one store per cycle, and all variants of \mpxshort{} checks \emph{increase} IPC.
In summary, the final IPC was \textasciitilde1.5--3 (compare to original IPC of 2), proving that the \mpxshort-protected program typically has \emph{approximately the same IPC as the original}.

As \figreftwo{perf}{instr} show, it causes major performance degradation.
It can be fixed, however; if the next generations of CPUs will provide the relative memory address calculation on other ports, the checks could be parallelized and performance will improve.
We speculate that GCC-\mpxshort{} could perform on par with AddressSanitizer in this case, because the instruction overheads are similar.
Accordingly, ICC version would be even better and the slowdowns might drop lower than 20\%.
But we must note that we do not have any hard proof for this speculation.

\subsection{\href{https://intel-mpx.github.io/design/\#operating-system}{Operating System}}
\label{sec:os}

The operating system has two main responsibilities in the context of \mpx: it handles bounds violations and manages BTs, i.e., creates and deletes them.
Both these actions are hooked to a new class of exceptions, \br, which has been introduced solely for \mpx{} and is similar to a page fault, although with extended functionality.

\myparagraph{Bounds exception handling}
If an \mpxshort-enabled CPU detects a bounds violation, i.e., if a referenced pointer appears to be outside of the checked bounds, \br{} is raised and the processor traps into the kernel (in case of Linux).
The kernel decodes the instruction to get the violating address and the violated bounds, and stores them in the \code{siginfo} structure.
Afterwards, it delivers the SIGSEGV signal to the application together with information about the violation in \code{siginfo}.
At this point the application developer has a choice: she can either provide an ad-hoc signal handler to recover or choose one of the default policies: crash, print an error and continue, or silently ignore it.

\myparagraph{Bounds tables management}
Two levels of bounds address translation are managed differently: BD is allocated only once by a runtime library (at application startup) and BTs have to be created dynamically on-demand.
The later is a task of OS.
The procedure is presented in \figref{bt-alloc}.
Each time an application tries to store pointer bounds \circled{1}, the CPU loads the corresponding entry from the BD and checks if it is a valid entry \circled{2}.
If the check fails, the CPU raises \br{} and traps into the kernel \circled{3}.
The kernel allocates a new BT \circled{4}, stores its address in the BD entry \circled{5} and returns in the user space \circled{6}.
Then, the CPU stores bounds in the newly created BT and continues executing the application in the normal mode of operation \circled{7}.

Since the application is oblivious of BT allocation, the OS also has to free these tables.
In Linux, this ``garbage collection'' is performed whenever a memory object is freed or, more precisely, unmapped.
OS goes through the elements of the object and removes all the corresponding BT entries.
If one of the tables becomes completely unused, OS will free the BT and remove its entry in the BD.

\myparagraph{\href{https://intel-mpx.github.io/microbenchmarks/\#os}{Microbenchmark}}
To illustrate the additional overhead of allocating and de-allocating BTs, we manufactured two microbenchmarks that showcase the worst case scenarios.
The first one stores a large set of pointers in such memory locations that each of them will have a separate BT, i.e., this benchmark indirectly creates a large number of bounds tables.
The second one does the same, but in addition, it frees all the memory right after it has been assigned, thus triggering BT de-allocation.
Our measurement results are shown in \tabref{os-micro} (note that we disabled all compiler optimizations to showcase the influence of OS alone).
In both cases, most of the runtime parameters (cache locality, branch misses, etc.) of the \mpxshort{}-protected version are equivalent to the native one.
However, the slowdown is noticeable---more than 2 times.
It is caused by a single parameter that varies---the number of instructions executed in the kernel space.
It means that the overhead is caused purely by the BT management in the kernel.
From this, we can conclude that OS itself can make an \mpxshort{}-protected application up to $2.3\times$~slower, although this scenario is quite rare.

% -*- root: ../main.tex -*-
\begin{table}

\footnotesize
\centering

\begin{tabular}{p{1.9cm} >{\centering\arraybackslash}p{1.2cm} >{\centering\arraybackslash}p{1.3cm} >{\centering\arraybackslash}p{1.3cm}}
%[-8pt]
%\hline
\noalign{\smallskip}
 &  & \multicolumn{2}{c}{\bfseries Increase in \# of instructions (\%)} \\
\cline{3-4}
\noalign{\vskip 1pt}
\bfseries Type & \bfseries Slowdown & \bfseries User space & \bfseries Kernel~space \\
 \noalign{\vskip -9pt}
\hline
\hline
\noalign{\vskip 1pt}
allocation                                   & $2.33\times$ & 7.5 & 160\\
%\hline
\\ [-9pt]
+ de-allocation                 & $2.25\times$ & 10  & 139\\
\hline
\end{tabular}

\caption{Worst-case OS impact on performance of MPX.}
\label{tab:os-micro}
%\vspace{3mm}
\end{table}

\smallskip

In this section, we discussed only Linux implementation.
However, all the same mechanisms can also be found in Windows.
The only significant difference is that \mpx{} support on Windows is done by a daemon, while on Linux the functionality is implemented in the kernel itself \cite{mpxenablingguide2016}. \remove

\subsection{\href{https://intel-mpx.github.io/design/\#compiler-and-runtime-library}{Compiler and Runtime Library}}
\label{sec:compiler}

Hardware \mpx{} support in the form of new instructions and registers significantly lowers performance overhead of each \emph{separate} bounds-checking operation.
However, the main burden of efficient, correct, and complete bounds checking of whole programs lies on the compiler and its associated runtime.

\myparagraph{\href{https://intel-mpx.github.io/design/\#compiler-support}{Compiler support}}
As of the date of this writing, only GCC 5.0+ and ICC 15.0+ compilers have support for \mpx{} \cite{gccmpx2016,mpxenablingguide2016}.
To enable \mpx{} protection of applications, both GCC and ICC introduce the new compiler pass called Pointer(s) Checker.
Enabling \mpx{} is intentionally as simple as adding a couple of flags to the usual compilation process:
\begin{lstlisting}[style=embedded]
  >>  gcc -fcheck-pointer-bounds -mmpx  test.c
  >>  icc -check-pointers-mpx=rw  test.c
\end{lstlisting}

In a glance, the Pointer Checker pass instruments the original program as follows. (1) It allocates static bounds for global variables and inserts \code{bndmk} instructions for stack-allocated ones. (2) It inserts \code{bndcl} and \code{bndcu} bounds-check instructions before each load or store from a pointer. (3) It moves bounds from one \code{bnd} register to another using \code{bndmov} whenever a new pointer is created from an old one. (4) It spills least used bounds to stack via \code{bndmov} if running out of available \code{bnd} registers. (5) It loads and stores the associated bounds via \code{bndldx} and \code{bndstx} respectively whenever a pointer is loaded/stored from/to memory.

One of the advantages of \mpx{}---in comparison to AddressSanitizer and SAFECode---is that it supports \emph{narrowing of struct bounds} by design.
Consider struct \code{obj} from \figref{mpx_example}.
It contains two fields: a 100B buffer \code{buf} and an integer \code{len} right after it.
It is easy to see that an off-by-one overflow in \code{obj.buf} will spillover and corrupt the adjacent \code{obj.len}.
AddressSanitizer and SAFECode by design cannot detect such intra-object overflows (though AddressSanitizer can be used to detect a subset of such errors \cite{asanintraobject2016}).
In contrast, \mpx{} can be instructed to narrow bounds when code accesses a specific field of a struct, e.g., on Line 10 in \figref{mpx_example}b.
Here, instead of checking against the bounds of the full object, the compiler would shrink \code{objptr\_b} to only four bytes and compare against these narrowed bounds on Lines 11--12.
Narrowing of bounds may require (sometimes intrusive) changes in the source code, and is enabled by default.

By default, the \mpxshort{} pass instruments both memory writes and reads: this ensures protection from buffer overwrites and buffer overreads.
The user can instruct the \mpxshort{} pass to instrument only writes.
The motivation is twofold.
First, instrumenting only writes significantly reduces performance overhead of \mpx{} (from $2.5\times$ to $1.3\times$ for GCC).
Second, the most dangerous bugs are those that overwrite memory (classic overflows to gain privileged access to the remote machine), and the only-writes protection can already provide sufficiently high security guarantees.

At least in GCC implementation, the pass can be fine-tuned via additional compilation flags.
In our experience, these flags provide no additional benefit in terms of performance, security, or usability.
For a full list of supported flags, refer to the official documentation of \mpx{} \cite{mpxenablingguide2016}.
\remove

For performance, compilers must try their best to optimize away redundant \mpxshort{} code.
There are two common optimizations used by GCC and ICC (also used, for example, in Baggy Bounds \cite{BaggyBounds2009}).
(1) Removing bounds-checks when the compiler can statically prove safety of memory access, e.g., access inside an array with a known offset.
(2) Moving (hoisting) bounds-checks out of simple loops.
Consider \figref{mpx_example}b.
If it is known that \code{M<=10}, then optimization (1) can remove always-true checks on Lines 6--7.
Otherwise, optimization (2) can kick in and move these checks before the loop body, saving two instructions on each iteration.

\myparagraph{\href{https://intel-mpx.github.io/design/\#runtime-library}{Runtime library}}
As a final step of the \mpxshort-enabled build process, the application must be linked against two \mpxshort-specific libraries: \code{libmpx} and \code{libmpxwrappers} (\code{libchkp} for ICC).

The \code{libmpx} library is responsible for \mpxshort{} initialization at program startup: it enables hardware and OS support and configures \mpxshort{} runtime options (passed through environment variables).
Most of these options concern debugging and logging, but two of them define security guarantees.
First, \code{CHKP\_RT\_MODE} must be set to ``stop'' in production use to stop the program immediately when a bounds violation is detected; set it to ``count'' only for debugging purposes.
Second, \code{CHKP\_RT\_BNDPRESERVE} defines whether application can call legacy, uninstrumented functions in external libraries; it must be enabled if the whole program is \mpxshort-protected.

By default, \code{libmpx} registers a signal handler that either halts execution or writes a debug message (depending on runtime options).
However, this default handler can be overwritten by the user's custom handler.
This can be useful if the program must shutdown gracefully or checkpoint its state.

Another interesting feature is that the user can instruct \code{libmpx} to disallow creation of BTs by the OS (see \secref{os}).
In this case, the \br{} exception will be forwarded directly to the program which can allocate BTs itself.
One scenario where this can come handy is when the user completely distrusts the OS, e.g., when using SGX enclaves \cite{SGXBounds2017}.

The \code{libmpxwrappers} library in GCC (and its analogue \code{libchkp} in ICC) contain wrappers for functions from C standard library (libc).
Similar to AddressSanitizer, \mpx{} implementations do not instrument libc and instead wrap all its functions with a bounds-checking counterparts.

% -*- root: ../main.tex -*-

\begin{table}
%\renewcommand{\arraystretch}{0.8}
%\vspace{-1mm}
\footnotesize
\centering

\begin{tabular}{p{5.0cm} p{0.7cm} p{0.7cm}}
%[-8pt]
%\hline
\multicolumn{1}{l}{\bfseries Compiler \& runtime issues}  & \multicolumn{1}{c}{\bfseries GCC} & \multicolumn{1}{c}{\bfseries ICC} \\
\hline
\hline
\noalign{\vskip 1pt}
-- Poor \mpxshort{} pass optimizations * & \multicolumn{1}{c}{22/38} & \multicolumn{1}{c}{3/38} \\
\\ [-8pt]
-- Bugs in \mpxshort{} compiler pass:  \\
% below bug is found in: x264, xalancbmk
\quad -- incorrect bounds during function calls & \multicolumn{1}{c}{--} & \multicolumn{1}{c}{2/38} \\
% below bug is found in: vips, milc, h264ref (for ICC 17)
\quad -- conflicts with auto-vectorization passes & \multicolumn{1}{c}{--} & \multicolumn{1}{c}{3/38} \\
% below bug is found in: dedup, blackscholes, swaptions
\quad -- corrupted stack due to C99 VLA arrays & \multicolumn{1}{c}{--} & \multicolumn{1}{c}{3/38} \\
% below bug is found in: xalancbmk (gcc-mpx-only-write)
\quad -- unknown internal compiler error & \multicolumn{1}{c}{1/38} & \multicolumn{1}{c}{--} \\
\\ [-8pt]
-- Bugs and issues in runtime libraries: \\
\quad -- Missing wrappers for libc functions & \multicolumn{1}{c}{all} & \multicolumn{1}{c}{all} \\
% below bug is found in: RIPE memcpy issues
\quad -- Nullified bounds in \code{memcpy} wrapper  & \multicolumn{1}{c}{all} & \multicolumn{1}{c}{--} \\
% below bug is found in: nginx under SGX environment
%\quad -- Sanity-check bug in \code{memcpy} wrapper  & nginx & \multicolumn{1}{c}{--} \\
% below bug is found in: wordcount (and probably all others)
\quad -- Performance bug in \code{memcpy} wrapper   & \multicolumn{1}{c}{--} & \multicolumn{1}{c}{all} \\
\\ [-8pt]
\hline
\\ [-8pt]
\multicolumn{3}{r}{\scriptsize{*One compiler has $>10\%$ worse results than the other}}
\\
\end{tabular}
\vspace{-1mm}
\caption{Issues in the compiler pass and runtime libraries of \mpx. Columns 2 and 3 show number of affected programs (out of total 38).\protect\footnotemark}
\label{tab:compiler}
%\vspace{1mm}
\end{table}
\footnotetext{All bugs were acknowledged by developers. Bug reports:\\
	\tiny \url{https://software.intel.com/en-us/forums/intel-c-compiler/topic/700550};
	\url{https://software.intel.com/en-us/forums/intel-c-compiler/topic/700675};
	\url{https://software.intel.com/en-us/forums/intel-c-compiler/topic/701764};
	\url{https://gcc.gnu.org/bugzilla/show_bug.cgi?id=78631}
%	\url{https://github.com/gcc-mirror/gcc/commit/4b27618793877baa48eba39fa0b9c0dc209cf19a}
}

\myparagraph{Issues}
For both GCC and ICC, the compiler and runtime support have a number of issues summarized in \tabref{compiler}.

Concerning performance, current implementations of GCC and ICC take different stances when it comes to optimizing \mpxshort{} code.
GCC is conservative and prefers stability of original programs over performance gains.
On many occasions, we noticed that the GCC \mpxshort{} pass \emph{disables} other optimizations, e.g., loop unrolling and autovectorization.
It also hoists bounds-checks out of loops less often than ICC does.
ICC, on the other hand, is more aggressive in its \mpxshort{}-related optimizations and does \emph{not} prevent other aggressive optimizations from being applied.
Unfortunately, this intrusive behavior renders ICC's pass less stable: we detected three kinds of compiler bugs due to incorrect optimizations.

We also observed issues with the runtime wrapper libraries.
First, only a handful of most widely-used libc functions are covered, e.g., \code{malloc}, \code{memcpy}, \code{strlen}, etc.
This leads to undetected bugs when other functions are called, e.g., the bug with \code{recv} in \secref{nginx}.
For use in production, these libraries must be expanded to cover \emph{all} of libc.
Second, while most wrappers follow a simple pattern of ``check bounds and call real function'', there exist more complicated cases.
For example, \code{memcpy} must be implemented so that it copies not only the contents of one memory area to another, but also all associated pointer bounds in BTs.
GCC library uses a fast algorithm to achieve this, but ICC's \code{libchkp} has a performance bottleneck (see also \secref{study}).

\begin{figure}[t]
	%	\vspace{2mm}
	\centering
	\href{https://intel-mpx.github.io/microbenchmarks/#performance}{\includegraphics[scale=0.6]{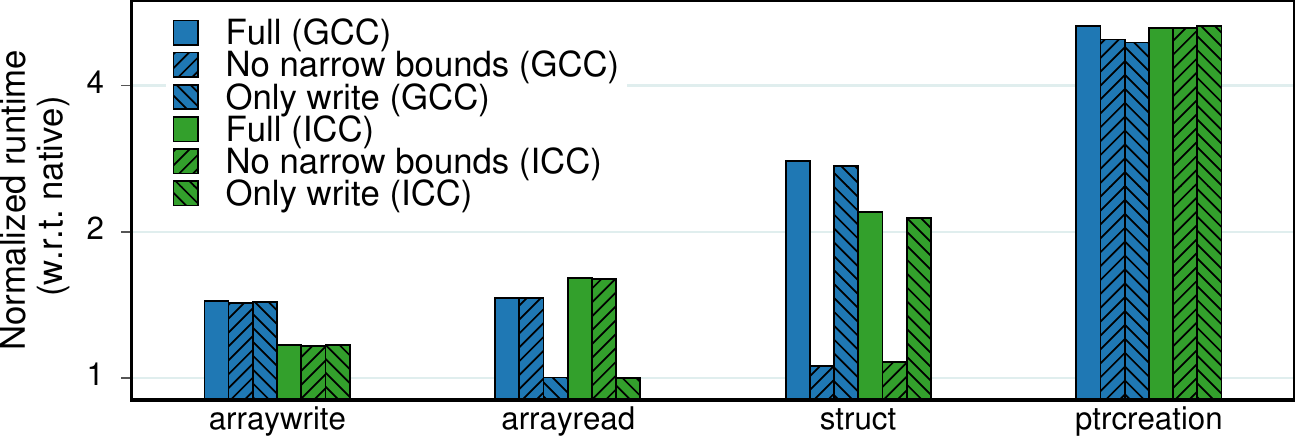}}
	\caption{\mpx{} overheads in 3 possible scenarios: application is dominated by bounds-checking (\emph{arraywrite} and \emph{arrayread}), by bounds creation and narrowing (\emph{struct}), and by bounds propagation (\emph{ptrcreation}).}
	\label{fig:micro_perf}
	%	\vspace{2mm}
\end{figure}

\myparagraph{\href{https://intel-mpx.github.io/microbenchmarks/\#performance}{Microbenchmarks}}
To understand the impact of different compiler flags and optimizations, we wrote four microbenchmarks, each highlighting a separate \mpxshort{} feature.
Two benchmarks---\emph{arraywrite} and \emph{arrayread}---perform writes to/reads from memory and stress \code{bndcl} and \code{bndcu} accordingly.
The \emph{struct} benchmark writes in an inner array inside a struct and stresses the bounds-narrowing feature via \code{bndmk} and \code{bndmov}.
Finally, the \emph{ptrcreation} benchmark constantly assigns new values to pointers and stresses bounds propagation via \code{bndstx}.
\figref{micro_perf} shows the performance overheads over native versions.

We can notice several interesting details.
First, \emph{arraywrite} and \emph{arrayread} represent bare overhead of bounds-checking instructions (all in registers), $50\%$ in this case.
\emph{struct} has a higher overhead of $2.1-2.8\times$ due to the more expensive making and moving of bounds to and from the stack.
The $5\times$ overhead of \emph{ptrcreation} is due to storing of bounds---the most expensive \mpxshort{} operation (see \secref{hardware}).
Such high overhead is alarming because pointer-intensive applications require many loads and stores of bounds.

Second, there is a $25\%$ difference between GCC and ICC in \emph{arraywrite}.
This is the effect of optimizations: GCC's \mpxshort{} pass blocks loop unrolling while ICC's implementation takes advantage of it.
(Interestingly, the same happened in case of \emph{arrayread} but the native ICC version was optimized even better, which led to a relatively poor performance of ICC's \mpxshort.)

Third, the overhead of \emph{arrayread} becomes negligible with the only-writes \mpxshort{} version: the only memory accesses in this benchmark are reads which are left uninstrumented.
Finally, the same logic applies to \emph{struct}---disabling narrowing of bounds effectively removes expensive \code{bndmk} and \code{bndmov} instructions and lowers performance overhead to a bare minimum.

\subsection{\href{https://intel-mpx.github.io/design/\#application}{Application}}
\label{sec:application}

At the application level, we observed two main issues of \mpx{}.
First, \mpx{} cannot support several widely-used C programming idioms (some by design, some due to implementation choices) and thus can break programs.
Second and more importantly, there is no support for multithreaded programs.

% -*- root: ../main.tex -*-

\begin{table}
%\renewcommand{\arraystretch}{0.8}
%\vspace{-1mm}
\footnotesize
\centering

\begin{tabular}{p{5.5cm} p{0.7cm} p{0.7cm}}
%[-8pt]
%\hline
\multicolumn{1}{l}{\bfseries Application-level issues}  & \multicolumn{1}{c}{\bfseries GCC} & \multicolumn{1}{c}{\bfseries ICC} \\
\hline                    
\hline
\noalign{\vskip 1pt}
% below bug is found in: ferret (+libjpeg), vips (+glib), x264, gcc, perlbench, xalancbmk
-- Flexible or variable-sized array (\code{arr[1]} / \code{arr[]}) & \multicolumn{1}{c}{7/38} & \multicolumn{1}{c}{7/38} \\ 
%\quad -- of them fixable: & \multicolumn{1}{c}{7/7} & \multicolumn{1}{c}{7/7} \\ 
\\ [-9pt]
% icc only: below bug is found in: ferret (+libjpeg), raytrace, xalancbmk
% gcc only: below bug is found in: xalancbmk
-- Accessing struct through struct field * & \multicolumn{1}{c}{1/38} & \multicolumn{1}{c}{3/38} \\ 
%\quad -- of them fixable: & \multicolumn{1}{c}{0/1} & \multicolumn{1}{c}{0/3} \\ 
\\ [-9pt]
% below bug is found in: gcc, soplex
-- Custom memory management & \multicolumn{1}{c}{2/38} & \multicolumn{1}{c}{2/38} \\ 
%\quad -- of them fixable: & \multicolumn{1}{c}{0/2} & \multicolumn{1}{c}{0/2} \\ 
\\ [-9pt]
\hline
\\ [-9pt]
\multicolumn{3}{r}{\scriptsize{* GCC affects less programs due to milder rules w.r.t. first field of struct}}
\\
\end{tabular}
\vspace{-1mm}
\caption{Applications may violate memory-model assumptions of \mpx. Columns 2 and 3 show number of misbehaving programs (out of total 38).}
\label{tab:application}
%\vspace{1mm}
\end{table}

\myparagraph{\href{https://intel-mpx.github.io/design/\#not-supported-c-idioms}{Not supported C idioms}}
As discussed previously, one of the main features of \mpx{}---narrowing of bounds---can increase security because the code that explicitly works with one field of a complex object will not be able corrupt other fields.
Unfortunately, our evaluation reveals that narrowing of bounds breaks many programs (see \tabref{application}).
The general problem is that C/C++ programs frequently deviate from the standard memory model \cite{BeyondPDP2015,DepthsOfC2016}.

A common C idiom (before C99) is flexible array fields with array size of one, e.g., \code{arr[1]}.
In practice, objects with such array fields have a dynamic size of \emph{more} than one item, but there is no way of \mpxshort{} knowing this at compile-time.
Thus, \mpx{} attempts to narrow bounds to one-item size whenever \code{arr} is accessed, which leads to false positives.
A similar idiom is variable-sized arrays also not supported by \mpx{}, e.g., \code{arr[]}.
These idioms are frequently seen in modern programs, see \tabref{application}, row 1.
Note that the C99-standard \code{arr[0]} is acceptable and does not break programs.

Another common idiom is using a struct field (usually the first field of struct) to access other fields of the struct.
Again, this breaks the assumptions of \mpx{} and leads to runtime \br{} exceptions (see \tabref{application}, row 2).
GCC makes an exception for this case since it is such a popular practice, but ICC is strict and does not have this special rule.

Finally, some programs introduce ``memory hacks'' for performance, ignoring restrictions of the C memory model completely.
The SPEC2006 suite has two such examples:
\emph{gcc} has its own complicated memory management with arbitrary type casts and in-pointer bit twiddling, and
\emph{soplex} features a scheme that moves objects from one memory region to another by adding an offset to each affected pointer (\tabref{application}, row 3).
%Some programs even introduce their own memory-management frameworks with drop-in replacements for \code{malloc} and \code{free}, like Apache with OpenSSL.
Both these cases lead to false positives.

Ultimately, all such non-compliant cases must be fixed (indeed, we \href{https://intel-mpx.github.io/dummy}{patched} flexible/variable-length array issues to work under \mpx).
However, sometimes the user may have strong incentives against modifying the original code.
In this case, she can opt for slightly worse security guarantees and disable narrowing of bounds via a \code{fno-chkp-narrow-bounds} flag.
Another non-intrusive alternative is to mark objects that must \emph{not} be narrowed (e.g., flexible arrays) with a special compiler attribute.

%GCC and ICC introduce \mpxshort-related intrinsics and compiler attributes which allow the user to have more fine-grained control over \mpx.
%In our experience, intrinsics are only meaningful in rare cases of writing wrappers for external libraries or for debugging.
%Attributes are more useful: the end user can disable/enable instrumentation of certain functions (e.g., functions with inline assembly) and mark objects that must \emph{not} be narrowed (e.g., flexible arrays).

\myparagraph{\href{https://intel-mpx.github.io/microbenchmarks/\#multithreading}{Multithreading issues}}
Current \mpx{} implementations may introduce false positives and negatives in multithreaded programs \cite{BeyondPDP2015}.
The problem arises because of the way \mpx{} loads and stores pointer bounds via its \code{bndldx} and \code{bndstx} instructions.
Recall from \secref{mpx} that whenever a pointer is loaded from main memory, its bounds must also be loaded from the corresponding bounds table (\figref{mpx_example}b, Lines 8-9).

% -*- root: ../main.tex -*-
%!TEX root = ../main.tex
\begin{figure}[t]
\begin{lstlisting}[frame=t,framesep=0pt,aboveskip=10pt,belowskip=0pt,numbers=none,label=algo:txexample]
  char* arr[1000]  |\hfill\color{Gray}{\itshape ~;; Array with MPX data race}|  
  char obj1  |\hfill\color{Gray}{\itshape ~;; Two adjacent objects $\rceil$}|
  char obj2  |\hfill\color{Gray}$\rfloor$|
\end{lstlisting}
\vspace{1mm}
\begin{lstlisting}[name=txexample-ir,frame=none,framesep=0pt,aboveskip=0pt,belowskip=0pt,numbersep=2pt,backgroundcolor=\color{Lavender}]
  while (true):  |\hfill\color{Gray}{\itshape ~;; Background thread}|
      for (i=0; i<1000; i++)   arr[i] = &obj1
      for (i=0; i<1000; i++)   arr[i] = &obj2  
\end{lstlisting}
\vspace{1mm}
\begin{lstlisting}[name=txexample-ir,frame=b,framesep=0pt,aboveskip=0pt,belowskip=0pt,numbersep=2pt]
  while (true):  |\hfill\color{Gray}{\itshape ~;; Main thread}|
      for (i=0; i<1000; i++)   *(arr[i] + offset)
\end{lstlisting}
\caption{This program breaks Intel MPX. If \code{offset=0} then MPX has false alarms, else --- undetected bugs.}
\label{fig:multithreading}
\vspace{3mm}
\end{figure}

Ideally, the load of the pointer and its bounds must be performed \emph{atomically} (same for stores).
However, nor the current hardware implementation neither GCC/ICC compilers enforce this atomicity.
This lack of proper multithreading support in \mpx{} can lead to (1) correct programs crashing due to false alarms, or (2) buggy programs being exploited \emph{even if} protected by \mpx.

Consider an example in \figref{multithreading}.
A ``pointer bounds'' data race happens on the \code{arr} array of pointers.
The background thread fills this array with all pointers to the first or to the second object alternately.
Meanwhile, the main thread accesses a whatever object is currently pointed-to by the array items.
Note that depending on the value of the constant \code{offset}, the original program is either always-correct or always-buggy: if \code{offset} is zero, then the main thread always accesses the correct object, otherwise it accesses an incorrect, adjacent object.
The second case, if found in a real code, introduces a vulnerability which could be exploited by an adversary.

With \mpx, additional \code{bndstx} instruction is inserted in Line 2 to store the bound corresponding to the first object (same for Line 3 and second object).
Also, a \code{bndldx} instruction is inserted in Line 5 to retrieve the bound for an object referenced by \code{arr[i]}.
Bound checks \code{bndcl} and \code{bndcu} are also added at Line 5, before the actual access to the object.
Now, the following race can occur.
The main thread loads the pointer-to-first-object from the array and---right before loading the corresponding bound---is preempted by the background thread.
The background thread overwrites all array items such that they point to the second object, and also overwrites the corresponding bounds.
Finally, the main thread is scheduled back and loads the bound, however, the bound now corresponds to the second object.
The main thread is left with the pointer to the first object but with the bounds of the second one.

We implemented this test case in C and compiled it with both GCC and ICC.
As expected, the \mpxshort-enabled program had both false positives and false negatives.

In case of a correct original program (i.e., with \code{offset=0}), such discrepancy leads to a \emph{false positive} when actually accessing the object at Line 5.
Indeed, the pointer to the object is correct but the bounds were overwritten by the background thread, so \mpxshort{} triggers a false-alarm exception.
Debugging the root cause of such non-deterministic pseudo-bugs would be a frustrating experience for end users.

The case of an originally buggy program (with \code{offset=1}) is more disconcerting.
After all, \mpx{} is supposed to detect all out-of-bounds accesses, but in this example \mpx{} introduces \emph{false negatives}!
Here, the pointer to the first object plus offset incorrectly lends into the second object.
But since the main thread checks against the bounds of the second object, this bug is not caught by \mpx.
We believe that this implementation flaw---that out-of-bounds bugs can \emph{sometimes} go unnoticed---can scare off users of multithreaded applications.
We also believe that a resourceful hacker would be able to construct an exploit that, based on these findings, could overcome \mpx{} defense with a high probability \cite{ConcurrencyAttacks2012}.

We must note however that we did not observe incorrect behavior in Phoenix and PARSEC multithreaded benchmark suites---we were lucky not to encounter programs that break \mpx.

%  (e.g., protecting such snippets via locks or utilizing hardware transactional memory)

For safe use in multithreaded programs, \mpxshort{} instrumentation must enforce atomicity of loading/storing pointers and their bounds.
At the software (compiler) level, this dictates the use of some synchronization primitive around each pair of \code{mov-bndldx/bndstx}, being it fine-grained locks, hardware transactional memory, or atomics.
Whatever primitive is chosen, we conjecture a significant drop in performance of \mpx.

A solution at a microarchitectural level would be to merge the pairs \code{mov-bndldx/bndstx} and assure their atomic execution.
The instruction decoder could detect a \code{bndldx}, find the corresponding pointer \code{mov} in the instruction queue, and instruct the rest of execution to handle these instructions atomically.
However, we believe this solution could require intrusive changes to the CPU front-end.
Moreover, it would significantly limit compiler optimization capabilities.

% \input{3_methodology}
% -*- root: main.tex -*-
%!TEX root = main.tex
\vspace{-2mm}
\section{Measurement Study}
\label{sec:study}

In this section we answer the following questions:

\begin{itemize}
    \item What is the performance penalty of \mpx{}?
    \begin{itemize}
        \item How much slower does a program become?
        \item How does memory consumption change?
        \item How does protection affect scalability of multithreaded programs?
    \end{itemize}
    \item What level of security does \mpx{} provide?
    \item What usability issues arise when \mpx{} is applied?
\end{itemize}

% For the sake of conciseness, the testbed and experiment methodology descriptions are skipped; they can be found on \href{https://intel-mpx.github.io/methodology/}{our website page}.

\subsection{\href{https://intel-mpx.github.io/methodology/}{Experimental Setup}}

All the experimental infrastructure was build using Fex \cite{fex} benchmarking framework with corresponding changes for the required build types, measurement tools, and for certain experimental procedures.

\myparagraph{Testbed}
All the experiments were performed on the following setup:

\begin{enumerate}
    \item \emph{Hardware:}
    \begin{itemize}
        \item Intel(R) Xeon(R) CPU E3-1230 v5 @ 3.40GHz
        \item 1 socket, 8 hyper-threads, 4 physical cores
        \item CPU caches: \mbox{L1d = 32KB}, \mbox{L1i = 32KB}, \mbox{L2 = 256KB}, shared \mbox{L3 = 8MB}
        \item 64 GB of memory
    \end{itemize}
    \item \emph{Network.} For experiments on case studies, we used two machines with the network bandwidth between them equal to 938 Mbits/sec as measured by iperf.
    \item \emph{Software infrastructure:}
    \begin{itemize}
        \item Kernel: 4.4.0
        \item GLibC: 2.21
        \item Binutils: 2.26.1
    \end{itemize}
    \item \emph{Compilers:}
    \begin{itemize}
        \item GCC 6.1.0. Configured with:
\begin{lstlisting}[style=embedded]
--enable-languages=c,c++ --enable-libmpx
--enable-multilib --with-system-zlib
\end{lstlisting}
        \item ICC 17.0.0
        \item Clang/LLVM 3.8.0 (AddressSanitizer). Configured with:
\begin{lstlisting}[style=embedded]
-G "Unix Makefiles"
-DCMAKE_BUILD_TYPE="Release"
-DLLVM_TARGETS_TO_BUILD="X86"
\end{lstlisting}
        \item Clang/LLVM 3.2.0 (SAFECode). Configured with:
\begin{lstlisting}[style=embedded]
-G "Unix Makefiles"
-DCMAKE_BUILD_TYPE="Release"
-DLLVM_TARGETS_TO_BUILD="X86"
\end{lstlisting}
        \item Clang/LLVM 3.4.0 (SoftBound). Configured with:
\begin{lstlisting}[style=embedded]
--enable-optimized --disable-bindings
\end{lstlisting}
    \end{itemize}
\end{enumerate}

\myparagraph{Measurement tools}
We used the following tools for measurements:

\begin{itemize}
    \item \emph{perf stat}. Our main tool used to measure all CPU-related parameters.
    The full list includes:
\begin{lstlisting}[style=embedded,language=C]
-e cycles,instructions,instructions:u,instructions:k
-e branch-instructions,branch-misses
-e dTLB-loads,dTLB-load-misses
-edTLB-stores,dTLB-store-misses
-e L1-dcache-loads,L1-dcache-load-misses
-e L1-dcache-stores,L1-dcache-store-misses
-e LLC-loads,LLC-load-misses
-e LLC-store-misses,LLC-stores
\end{lstlisting}
    Not to introduce additional measurement error, we measured these parameters in parts, 8 parameters at a time.
    \item \emph{time}. Since perf does not provide capabilities for measuring physical memory consumption of a process, we used time \code{--verbose} and collected maximum resident set size.
    \item \emph{Intel Pin}. To gather \mpx{} instruction statistics, we developed a Pin tool.
    Full code of our instrumentation can be found in the repository.
\end{itemize}

\myparagraph{Benchmarks}
We used three benchmark suits in our evaluation: PARSEC 3.0 \cite{Parsec}, Phoenix 2.0 \cite{Phoenix}, and SPEC CPU 2006 \cite{SPEC}.
To remove some of the previously found bugs, we applied \href{https://github.com/google/sanitizers/blob/master/address-sanitizer/spec/spec2006-asan.patch}{a patch to SPEC suite}.
Also, during our work, we found and fixed \href{https://intel-mpx.github.io/usability/\#changes}{a set of bugs} in them.

All the benchmarks were compiled together with the libraries they depend upon (except raytrace from PARSEC which requires X11 libraries).

\myparagraph{Build types}

\begin{itemize}
    \item \emph{GCC implementation of MPX.}
    \begin{itemize}
        \item Compiler flags:
\begin{lstlisting}[style=embedded]
-fcheck-pointer-bounds -mmpx
\end{lstlisting}
        \item Linker flags:
\begin{lstlisting}[style=embedded]
-lmpx -lmpxwrappers
\end{lstlisting}
        \item Environment variables:
\begin{lstlisting}[style=embedded]
CHKP_RT_BNDPRESERVE="0"
CHKP_RT_MODE="stop"
CHKP_RT_VERBOSE="0"
CHKP_RT_PRINT_SUMMARY="0"
\end{lstlisting}
        \item Subtypes:
        \begin{itemize}
            \item Disabled bounds narrowing:
\begin{lstlisting}[style=embedded]
-fno-chkp-narrow-bounds
\end{lstlisting}
            \item Protecting only memory writes, not reads:
\begin{lstlisting}[style=embedded]
-fno-chkp-check-read
\end{lstlisting}
        \end{itemize}
    \end{itemize}
    \item \emph{ICC implementation of MPX.}
    \begin{itemize}
        \item Compiler flags:
\begin{lstlisting}[style=embedded]
-check-pointers-mpx=rw
\end{lstlisting}
        \item Linker flags:
\begin{lstlisting}[style=embedded]
-lmpx
\end{lstlisting}
        \item Environment variables:
\begin{lstlisting}[style=embedded]
CHKP_RT_BNDPRESERVE="0"
CHKP_RT_MODE="stop"
CHKP_RT_VERBOSE="0"
CHKP_RT_PRINT_SUMMARY="0"
\end{lstlisting}
        \item Subtypes:
        \begin{itemize}
            \item Disabled bounds narrowing:
\begin{lstlisting}[style=embedded]
-no-check-pointers-narrowing
\end{lstlisting}
            \item protecting only memory writes, not reads:
\begin{lstlisting}[style=embedded]
-check-pointers-mpx=write
// instead of
-check-pointers-mpx=rw
\end{lstlisting}
        \end{itemize}
    \end{itemize}
    \item \emph{AddressSanitizer (both GCC and Clang).}
    \begin{itemize}
        \item Compiler flags:
\begin{lstlisting}[style=embedded]
-fsanitize=address
\end{lstlisting}
        \item Environment variables:
\begin{lstlisting}[style=embedded]
ASAN_OPTIONS="verbosity=0:\
detect_leaks=false:\
print_summary=true:\
halt_on_error=true:\
poison_heap=true:\
alloc_dealloc_mismatch=0:\
new_delete_type_mismatch=0"
\end{lstlisting}
        \item Subtypes:
        \begin{itemize}
            \item Protecting only memory writes, not reads:
\begin{lstlisting}[style=embedded]
--param asan-instrument-reads=0
\end{lstlisting}
        \end{itemize}
    \end{itemize}
    \item \emph{SoftBound.}
    \begin{itemize}
        \item Compiler flags:
\begin{lstlisting}[style=embedded]
-fsoftboundcets -flto -fno-vectorize
\end{lstlisting}
        \item Linker flags:
\begin{lstlisting}[style=embedded]
-lm -lrt
\end{lstlisting}
    \end{itemize}
    \item \emph{SAFECode.}
    \begin{itemize}
        \item Compiler flags:
\begin{lstlisting}[style=embedded]
-fmemsafety -g
-fmemsafety-terminate -stack-protector=1
\end{lstlisting}
    \end{itemize}
\end{itemize}

\myparagraph{Experiments}
Each program was executed 10 times, and the results were averaged using arithmetic mean (note, we made sure that variance is very low and it is safe to use arithmetic mean).
The mean across different programs in the benchmark suite was calculated using geometric mean.
Geometric mean was also used to calculate the ``final'' mean across three benchmark suites.

We performed the following types of experiments:

\begin{itemize}
    \item normal: experiments on a single thread (serialized) and with fixed input
    \item multithreaded: experiments on 2, 4, and 8 threads
    \item variable inputs: experiments with increasing input size (5 runs, each next one with an input twice bigger than the previous)
\end{itemize}

The results were checked to fulfill the following criteria:

\begin{itemize}
    \item application compiled successfully
    \item application run successfully (with zero exit code)
    \item the output is equal to the output of non-protected application (if it is deterministic)
\end{itemize}

Values of coefficient of variation (CV) are presented in \tabref{cov}.

% -*- root: ../main.tex -*-
\begin{table}

\footnotesize
\centering

\begin{tabular}{l p{2.5cm} p{3cm}}
%[-8pt]
%\hline
\bfseries Experiment & \bfseries Average CV, \% & \bfseries Maximum CV, \% \\
\hline
\hline
\noalign{\vskip 1pt}
Phoenix					& 0.34 & 3.87 \\
%\hline
PARSEC					& 0.28 & 3.75 \\
%\hline
% \\ [-6pt]
SPEC					& 0.41 & 3.96 \\
%\hline
% \\ [-6pt]
\bfseries All			& \bfseries 0.35 & \bfseries 3.96 \\
%\hline

\hline
\end{tabular}

\caption{Variation of results in our experiments.}
\label{tab:cov}
%\vspace{3mm}
\end{table}

\subsection{\href{https://intel-mpx.github.io/performance/}{Performance}}
\label{sec:performance}

% To evaluate overheads incurred by \mpx{}, we tested three benchmark suits: Phoenix 2.0 \cite{Phoenix}, PARSEC 3.0 \cite{Parsec} and SPEC CPU2006 \cite{SPEC}.
To evaluate overheads incurred by \mpx{}, we tested it on three benchmark suites: Phoenix 2.0 \cite{Phoenix}, PARSEC 3.0 \cite{Parsec}, and SPEC CPU2006 \cite{SPEC}.
To put the results into context, we measured not only the ICC and GCC implementations of \mpx{}, but also AddressSanitizer, SAFECode, and SoftBound (see \secref{back} for details).

\begin{figure*}[t]
    \centering
    \href{https://intel-mpx.github.io/performance/#performance}{\includegraphics[scale=0.66]{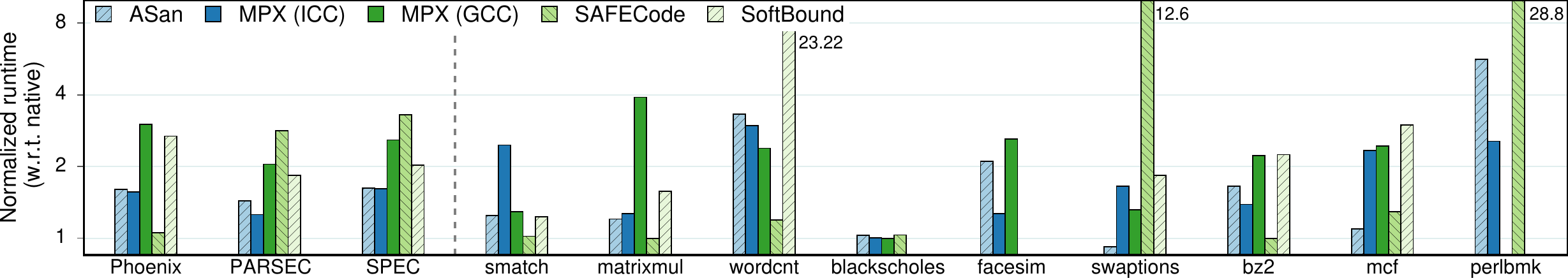}}
    \vspace{-16pt}
    \caption{Performance (runtime) overhead with respect to native version. (Lower is better.)}
    \label{fig:perf}
    \vspace{2pt}

    \centering
    \href{https://intel-mpx.github.io/performance/#instruction-overhead}{\includegraphics[scale=0.66]{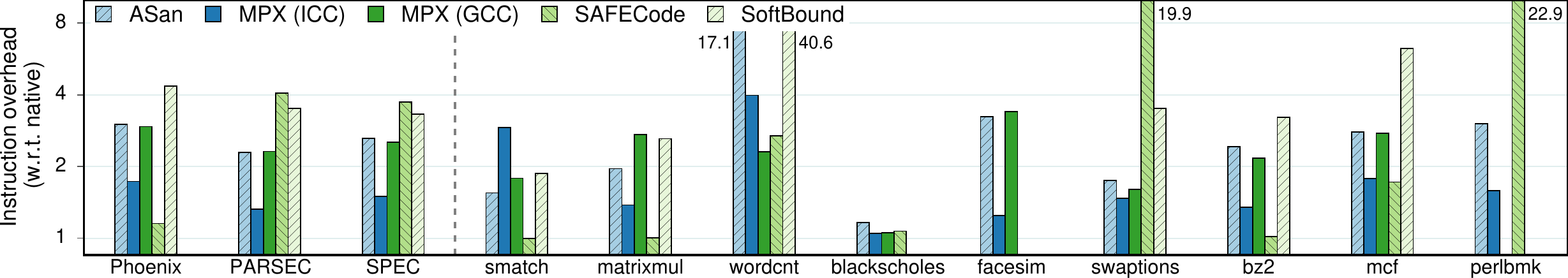}}
    \vspace{-16pt}
    \caption{Increase in number of instructions with respect to native version. (Lower is better.)}
    \label{fig:instr}
    \vspace{2pt}

    \centering
    \href{https://intel-mpx.github.io/performance/#ipc}{\includegraphics[scale=0.66]{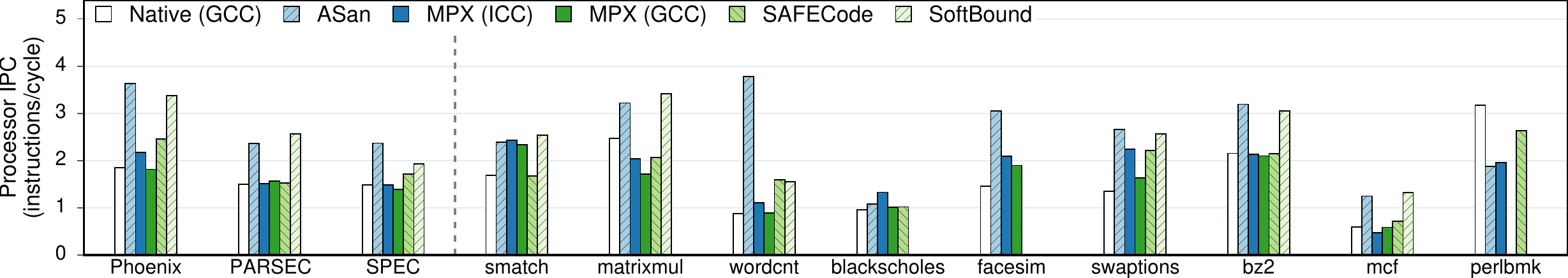}}
    \vspace{-16pt}
    \caption{IPC (instructions/cycle) numbers for native and protected versions. (Higher is better.)}
    \label{fig:ipc}
    \vspace{-6pt}
\end{figure*}

\myparagraph{\href{https://intel-mpx.github.io/performance/\#performance}{Runtime overhead}}
We start with the single most important parameter: runtime overhead (see \figref{perf}).

\begin{figure*}[t]
    \centering
    \href{https://intel-mpx.github.io/performance/#cache-utilization}{\includegraphics[scale=0.568]{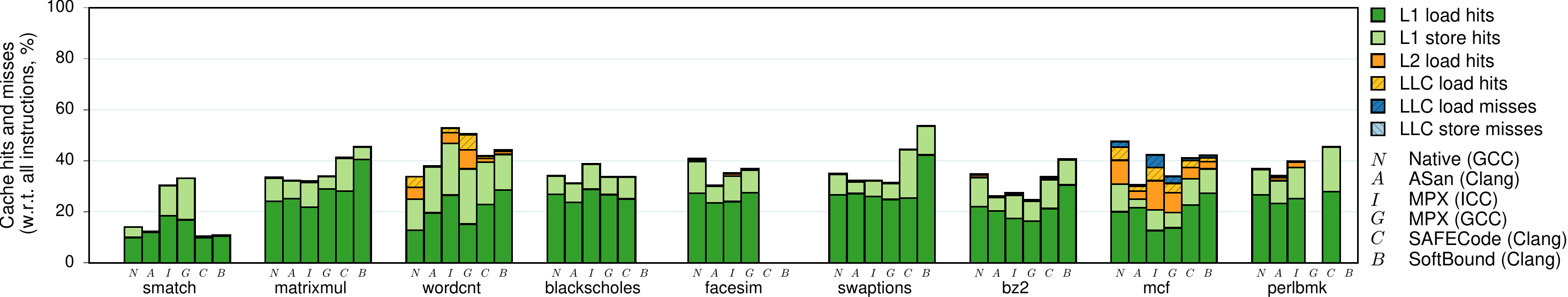}}
    \vspace{-16pt}
    \caption{CPU cache behavior of native and protected versions.}
    \label{fig:cache}
    \vspace{-6pt}
\end{figure*}

\begin{figure*}[t]
    \centering
    \href{https://intel-mpx.github.io/performance/#mpx-instructions}{\includegraphics[scale=0.58]{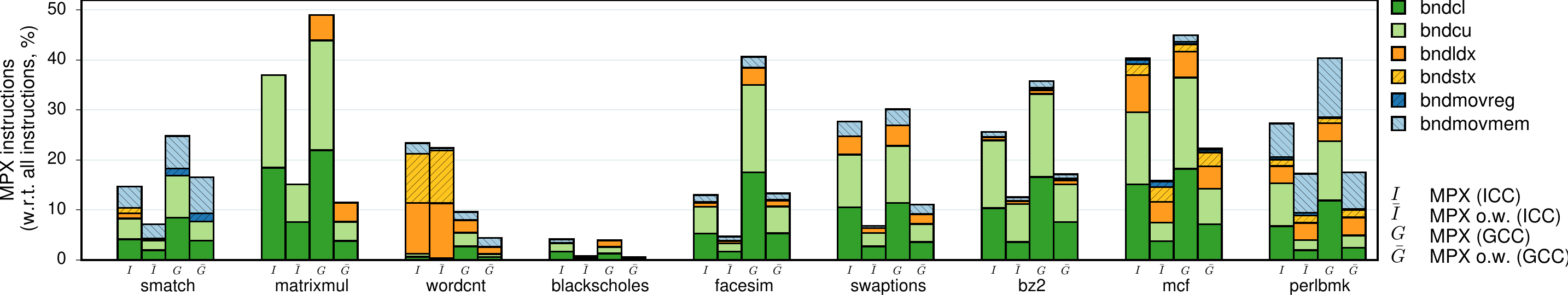}}
    \vspace{-16pt}
    \caption{Shares of \mpx{} instructions with respect to all executed instructions. (Lower is better.)}
    \label{fig:mpxcount}
    \vspace{-6pt}
\end{figure*}

First, we note that ICC-\mpxshort{} performs significantly better than GCC-\mpxshort{}.
At the same time, ICC is less usable: only 30 programs out of total 38 (79\%) build and run correctly, whereas 33 programs out of 38 (87\%) work under GCC (see also \secref{usability}).

AddressSanitizer, despite being a software-only approach, performs on par with ICC-\mpxshort{} and better than GCC-\mpxshort{}.
This unexpected result testifies that the hardware-assisted performance improvements of \mpx{} are offset by its complicated design and suboptimal instructions.
Although, AddressSanitizer provides worse security guarantees than \mpx{} (\secref{security}).

% Why is it here? To have more sparce placement of figures.
\begin{figure*}[t]
	\centering
	\href{https://intel-mpx.github.io/performance/#memory-consumption}{\includegraphics[scale=0.66]{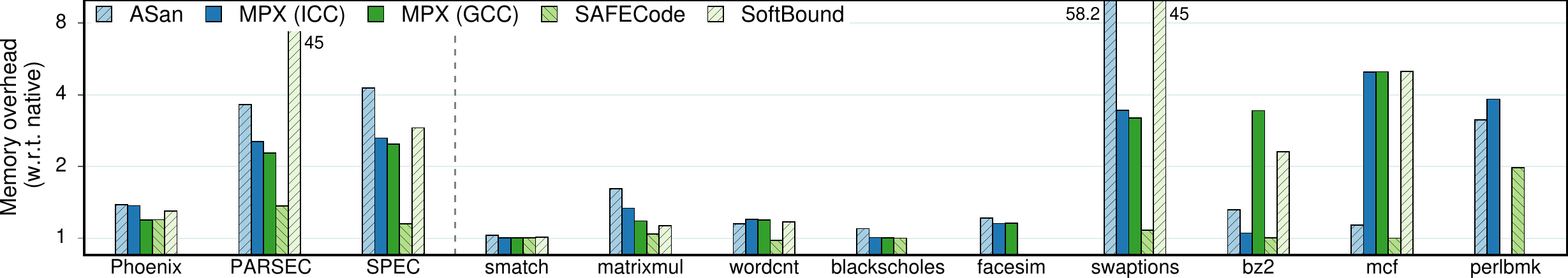}}
	\vspace{-16pt}
	\caption{Memory overhead with respect to native version. (Lower is better.)}
	\vspace{-6pt}
	\label{fig:mem}
\end{figure*}

SAFECode and SoftBound show good results on Phoenix programs, but behave much worse---both in terms of performance and usability---on PARSEC and SPEC.
First, consider SAFECode on Phoenix: due to the almost-pointerless design and simplicity of Phoenix programs, SAFECode achieves a low overhead of 5\%.
However, it could run only 18 programs out of 31 (58\%) on PARSEC and SPEC and exhibited the highest overall overheads.
SoftBound executed only 7 programs on PARSEC and SPEC (23\%).
Moreover, both SAFECode and SoftBound showed unstable behavior: some programs had overheads of more than $20\times{}$.

\myparagraph{\href{https://intel-mpx.github.io/performance/\#instruction-overhead}{Instruction overhead}}
In most cases, performance overheads are dominated by a single factor: the increase in number of instructions executed in a protected application.
It can be seen if we compare \figreftwo{perf}{instr}; there is a strong correlation between them.

As expected, the optimized \mpxshort{} (i.e., ICC version) has low instruction overhead due to its HW assistance (\textasciitilde70\% lower than AddressSanitizer).
Thus, one could expect sufficiently low performance overheads of \mpx{} once the throughput and latencies of \mpx{} instructions improve (see \secref{hardware}).

Instruction overhead of \mpx{} may also come from the management of Bounds Tables (see \secref{os}).
Our microbenchmarks show that it can cause a slowdown of more than 100\% in the worst case.
However, we did not observe a noticeable impact in real-world applications.
Even those applications that create hundreds of BTs exhibit a minor slowdown in comparison to other factors.

\myparagraph{\href{https://intel-mpx.github.io/performance/\#ipc}{IPC}}
Many programs do not utilize the CPU execution-unit resources fully.
For example, the theoretical IPC (instructions/cycle) of our machine is \textasciitilde5, but many programs achieve only 1--2 IPC in native executions (see \figref{ipc}).
Thus, memory-safety techniques benefit from underutilized CPU and partially mask their performance overhead.

The most important observation here is that \mpx{} does not increase IPC.
Our microbenchmarks (\secref{hardware}) indicate that this is caused by contention of \mpxshort{} bounds-checking instructions on one execution port.
If this functionality would be available on more ports, \mpx{} would be able to use instruction parallelism to a higher extent and the overheads would be lower.
%It is caused by the fact that bounds checking instructions (\code{bndcl}, \code{bndcu}, and \code{bndcn}) are considered by the processor as a data dependency to subsequent memory accesses.

At the same time, software-only approaches---especially AddressSanitizer and SoftBound---significantly increase IPC, partially hiding their performance overheads.

\begin{figure*}[t]
	\centering
	\href{https://intel-mpx.github.io/performance/#performance-1}{\includegraphics[scale=0.66]{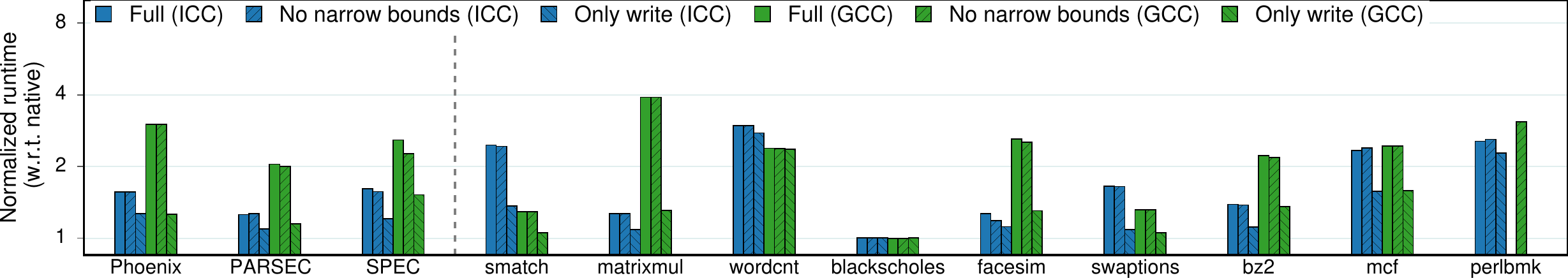}}
	\vspace{-16pt}
	\caption{Impact of \mpxshort{} features---narrowing and only-write protection---on performance. (Lower is better.)}
	\vspace{-6pt}
	\label{fig:mpx_feature_mem}
\end{figure*}

\begin{figure*}[t]
	\centering
	\href{https://intel-mpx.github.io/performance/#memory-consumption-1}{\includegraphics[scale=0.66]{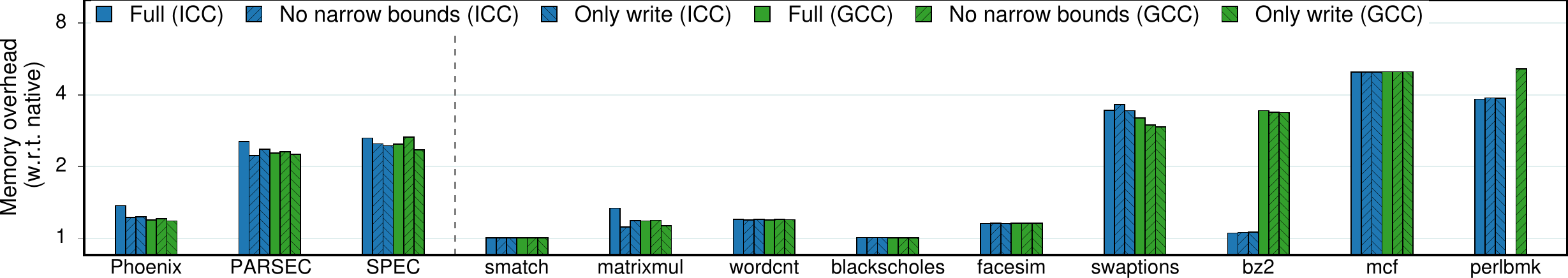}}
	\vspace{-16pt}
	\caption{Impact of \mpxshort{} features---narrowing and only-write protection---on memory. (Lower is better.)}
	\vspace{-6pt}
	\label{fig:mpx_feature_perf}
\end{figure*}

\begin{figure*}[t]
	\centering
	\href{https://intel-mpx.github.io/performance/#multithreading}{\includegraphics[scale=0.66]{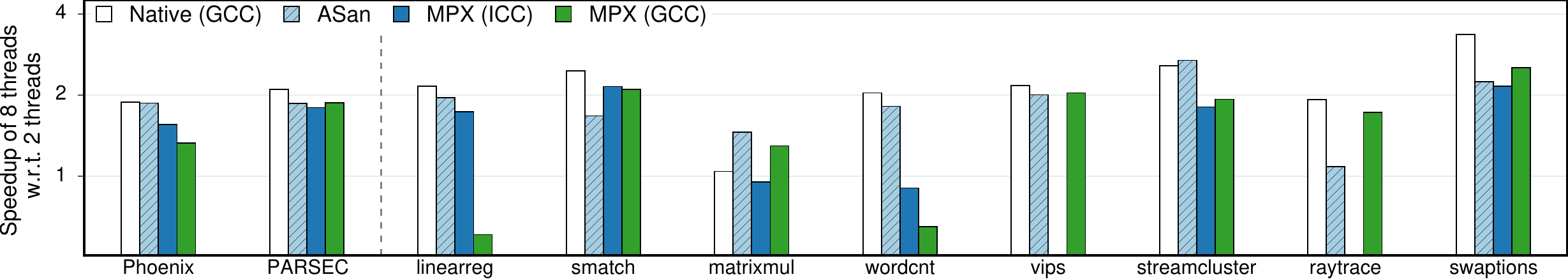}}
	\vspace{-16pt}
	\caption{Relative speedup (scalability) with 8 threads compared to 2 threads. (Higher is better.)}
	\vspace{-6pt}
	\label{fig:multi}
\end{figure*}

\myparagraph{\href{https://intel-mpx.github.io/performance/\#cache-utilization}{Cache utilization}}
Some programs are memory-intensive and stress the CPU cache system.
If a native program has many L1 or LLC cache misses, then the memory subsystem becomes the bottleneck.
In these cases, memory-safety techniques can partially hide their performance overhead.

It can be illustrated with the \emph{wordcnt} example compiled with ICC-\mpxshort{} (\figref{cache}).
It has a huge instruction overhead of $4\times$, IPC close to native, and (as we will see next) many expensive \code{bndldx} and \code{bndstx} operations.
And still its performance overhead is only $3\times$.
Why? It appears the native version of \emph{wordcnt} has a significant number of cache misses.
They have high performance cost and therefore can partially mask the overhead of ICC-\mpxshort{}.

\myparagraph{\href{https://intel-mpx.github.io/performance/\#mpx-instructions}{\mpx{} instructions}}
In the case of \mpx{}, one of the most important performance factors is the type of instructions that are used in instrumentation.
In particular, storing (\code{bndstx}) and loading (\code{bndldx}) bounds require two-level address translation---a very expensive operation that can break cache locality (see \secref{hardware}).
To prove it, we measured the shares of \mpxshort{} instructions in the total number of instructions of each program (\figref{mpxcount}).

As expected, a lion share of all \mpxshort{} instructions are bounds-checking \code{bndcl} and \code{bndcu}.
Additionally, many programs need \code{bndmov} to move bounds from one register to another (\code{bndmovreg}) or spill bounds on stack (\code{bndmovmem}).
Finally, pointer-intensive programs require the use of expensive \code{bndstx} and \code{bndldx} to store/load bounds in Bounds Tables.

There is a strong correlation between the share of \code{bndstx} and \code{bndldx} instructions and performance overheads.
For example, \emph{matrixmul} under ICC-\mpxshort{} almost exclusively contains bounds checks: accordingly, there is a direct mapping between instruction and performance overheads.
However, the GCC-\mpxshort{} version is less optimized and inserts many \code{bndldx}s, which leads to a significantly higher performance overhead.

The ICC-\mpxshort{} version of \emph{wordcnt} has a ridiculous share of \code{bndldx}/\code{bndstx} instructions.
This is due to a performance bug in libchkp library of ICC that uses a naive algorithm for the \code{memcpy} wrapper (see \secref{compiler}).

\myparagraph{\href{https://intel-mpx.github.io/performance/\#memory-consumption}{Memory consumption}}
In some scenarios, memory overheads (more specifically, resident set size overheads) can be a limiting factor, e.g., for servers in data centers which co-locate programs and perform frequent migrations.
Thus, \figref{mem} shows memory overhead measurements.

On average, \mpx{} has a $2.1\times$ memory overhead under ICC version and $1.9\times$ under GCC.
It is a significant improvement over AddressSanitizer ($2.8\times$).
There are three main reasons for that.
First, AddressSanitizer changes memory layout of allocated objects by adding ``redzones'' around each object.
Second, it maintains a ``shadow zone'' that is directly mapped to main memory and grows linearly with the program's working set size.
Third, AddressSanitizer has a ``quarantine'' feature that restricts the reuse of freed memory
\footnote{Quarantine is a temporal-protection feature and, in principle, it gives an unfair advantage to \mpx{} which lacks this kind of protection.
Indeed, if quarantine zone is disabled, AddressSanitizer's memory overhead drops on average to \textasciitilde$1.5\times$ for both PARSEC and SPEC, although the performance overhead is not influenced.
We did not include this number into our main results because the goad of our study was to compare the solutions in their \emph{default} configuration, without any tweaks from the side of end user.}.
On the contrary, \mpx{} allocates space only for pointer-bounds metadata and has an intermediary Bounds Directory that trades lower memory consumption for longer access time.
Interestingly, SAFECode exhibits even lower memory overheads because of its pool-allocation technique.
Unfortunately, low memory consumption does not imply good performance.

\myparagraph{\href{https://intel-mpx.github.io/performance/\#mpx-features}{Influence of additional \mpx{} features}}
\mpx{} has two main features that influence both performance and security guaranties (\secref{compiler}).
\emph{Bounds narrowing} increases security level but may harm performance.
\emph{Only-write protection}, on the other side, improves performance by disabling checks on memory reads.

The comparison of these features is presented in \figreftwo{mpx_feature_mem}{mpx_feature_perf}.
As we can see, bounds narrowing has a low impact on performance because it does not change the number of checks.
At the same time, it may slightly increase memory consumption because it has to keep more bounds.
Only-write checking has the opposite effect---having to instrument less code reduces the slowdown but barely has any impact on memory consumption.

\myparagraph{\href{https://intel-mpx.github.io/performance/\#multithreading}{Multithreading}}
To evaluate the influence of multithreading, we measured execution times of all benchmarks on 2 and 8 threads (see \figref{multi}).
%The approach for enabling multithreading was different for different benchmark suites: for Phoenix it was enough to set a corresponding compilation flag; Parsec required an alternative version of the source code (supplied with the suite).
Note that only Phoenix and PARSEC are multithreaded (SPEC is not).
Also, both SoftBound and SAFECode are not thread-safe and therefore were excluded from measurements.
%\alex{This statement needs a proof}

As we can see from \figref{multi}, the difference in scalability is minimal.
For \mpx{}, it is caused by the absence of multithreading support, which means that no additional code is executed in multithreaded versions.
For AddressSanitizer, there is no need for explicit synchronization---the approach is thread-safe by design.

Peculiarly, GCC-\mpxshort{} experiences not speedups but slowdowns on \emph{linearreg} and \emph{wordcnt}.
Upon examining these cases, we found out that this anomaly is due to detrimental cache line sharing of BT entries.

%\emph{matrixmul} does not have a speedup in its native version.
%In a nutshell, there are $3.5\times$ more LLC-loads on 8 threads than on 2.
%This happens due to hyperthreading---our machine has 4 physical cores with L1 and L2 caches shared among each two threads.

%For \emph{raytrace}, AddressSanitizer seems to exhibit only small speedup when going from 2 threads to 8.
%In reality, this is not a problem of AddressSanitizer but of the Clang compiler itself.
%The plot shows the native GCC version which---a rare corner case---scales much better than the native Clang version ($2\times$ speedup in comparison to $1.1\times$).

For \emph{swaptions}, AddressSanitizer and \mpx{} scale significantly worse than native.
It turns out that these techniques do not have enough spare IPC resources to fully utilize 8 threads in comparison to the native version (the problem of hyperthreading).
Similarly, for \emph{streamcluster}, \mpx{} performs worse than AddressSanitizer and native versions.
This is again an issue with hyperthreading: \mpx{} instructions saturate IPC resources on 8 threads and thus cannot scale as good as native.

\myparagraph{\href{https://intel-mpx.github.io/performance/\#varying-input-sizes}{Varying inputs sizes}}
Different input sizes (working sets) may cause different cache behaviors, which in turn causes changes in overheads.
To investigate the extent of such effects, we ran several benchmarks with three inputs---small, medium, and large.
The results do not provide any unexpected insights and thus omitted from the paper (but can be found on the \href{https://intel-mpx.github.io/performance/#varying-input-sizes}{website}).
The general trend is that the input size has very little impact on performance overhead.

\subsection{\href{https://intel-mpx.github.io/security/}{Security}}
\label{sec:security}

\myparagraph{\href{https://intel-mpx.github.io/security/\#ripe}{RIPE testbed}}
We evaluated all approaches against the RIPE security testbed \cite{RIPE}.
RIPE is a synthesized C program that tries to attack itself in a number of ways, by overflowing a buffer allocated on the stack, heap, or in data or BSS segments.
RIPE can imitate up to 850 attacks, including shellcode, return-into-libc, and return-oriented programming.
%% compiler flags are: -O0 -g -fno-stack-protector -Wl,-z,execstack -U_FORTIFY_SOURCE -D_FORTIFY_SOURCE=0
%% ASLR was disabled via: sudo bash -c 'echo 0 > /proc/sys/kernel/randomize_va_space'
In our evaluation, even under relaxed security flags---we disabled Linux ASLR, stack canaries, and fortify-source and enabled executable stack---modern compilers were susceptible only to a small number of attacks.
Under native GCC, only 64 attacks survived, under ICC---34, and under Clang---38.\footnote{RIPE is specifically tailored to GCC, thus more attacks are possible under this compiler.}

The results for all approaches are presented in \tabref{ripe}.
Surprisingly, a default GCC-\mpxshort{} version showed very poor results, with 41 attacks (or $64\%$ of all possible attacks) succeeding.
As it turned out, the default GCC-\mpxshort{} flags are sub-optimal.
First, we found a bug in the \code{memcpy} wrapper which forced bounds registers to be nullified, so the bounds checks on \code{memcpy} were rendered useless (see \tabref{compiler}).
This bug disappears if the \code{BNDPRESERVE} environment variable is manually set to one.
Second, the \mpxshort{} pass in GCC does \emph{not} narrow bounds for the first field of a struct by default, in contrast to ICC which is more strict.
To catch intra-object overflows happening in the first field of structs--the case of RIPE code---one needs to pass the \code{-fchkp-first-field-has-own-bounds} flag to GCC.
When we enabled these two flags, all attacks were prevented; all next rows in the table were tested with these flags.

Other results are expected.
\mpx{} versions without narrowing of bounds overlook 14 intra-object overflow attacks, where a vulnerable buffer and a victim object live in the same struct.
The same attacks are overlooked by AddressSanitizer, SoftBound, and SAFECode.
Interestingly, AddressSanitizer has 12 working attacks, i.e., two attacks less than other approaches.
Though we did not inspect this in detail, AddressSanitizer was able to prevent two shellcode intra-object attacks on the heap.

We performed the same experiment with only-writes versions of these approaches, and the results were exactly the same.
This is explained by the fact that RIPE constructs only control-flow hijacking attacks and not information leaks (which could escape only-writes protection).

% -*- root: ../main.tex -*-
\begin{table}

\footnotesize
\centering

\begin{tabular}{l p{4.4cm}}
%[-8pt]
%\hline
\bfseries Approach & \bfseries Working attacks \\
\hline
\hline
\noalign{\vskip 1pt}
MPX (GCC) default *     & {\textbf{41/64} (all memcpy and intra-object of.)} \\
MPX (GCC)               & {\textbf{0/64} (none)} \\
MPX (GCC) no narrow     & {\textbf{14/64} (all intra-object overflows)} \\
\\ [-6pt]
MPX (ICC)               & \textbf{0/34} (none) \\
MPX (ICC) no narrow     & \textbf{14/34} (all intra-object overflows) \\
\\ [-6pt]
AddressSanitizer (GCC)  & \textbf{12/64} (all intra-object overflows) \\
%\\ [-6pt]
SoftBound (Clang)       & \textbf{14/38} (all intra-object overflows) \\
%\\ [-6pt]
SAFECode (Clang)        & \textbf{14/38} (all intra-object overflows) \\
\hline
\\ [-8pt]
\multicolumn{2}{r}{\scriptsize{*Without \code{-fchkp-first-field-has-own-bounds}}}
\\ [-2pt]
\multicolumn{2}{r}{\scriptsize{and with \code{BNDPRESERVE=0}}}
\\
\end{tabular}

\caption{Results of RIPE security benchmark. In Col. 2, ``41/64'' means that 64 attacks were successful in native GCC version, and 41 attacks remained in MPX version.}
\label{tab:ripe}
%\vspace{3mm}
\end{table}

\myparagraph{\href{https://intel-mpx.github.io/security/\#others}{Other detected bugs}}
During our experiments, we found 6 real out-of-bounds bugs (true positives).
Five of these bugs were already known, and one was detected by GCC-\mpxshort{} and was not previously reported.
%\dmitry{should we report??}

The bugs found are:
% https://github.com/OleksiiOleksenko/mpx_evaluation/wiki/ferret#error-2-fixed
(1) incorrect black-and-white input pictures leading to classic buffer overflow in \code{ferret},
% https://github.com/OleksiiOleksenko/mpx_evaluation/wiki/h264ref#h264ref
(2) wrong preincrement statement leading to classic off-by-one bug in \code{h264ref},
% https://github.com/OleksiiOleksenko/mpx_evaluation/wiki/perlbench#error-1-fixed
(3) out-of-bounds write in \code{perlbench},
% https://github.com/OleksiiOleksenko/mpx_evaluation/wiki/x264#error-2-wontfix
(4) benign intra-object buffer overwrite in \code{x264},
% https://github.com/OleksiiOleksenko/mpx_evaluation/wiki/h264ref#error-3-wontfix-real-bug
(5) benign intra-object buffer overread in \code{h264ref},
% https://github.com/OleksiiOleksenko/mpx_evaluation/wiki/perlbench#error-3-wontfix
and (6) intra-object buffer overwrite in \code{perlbench}.

All of these bugs were detected by GCC-\mpxshort{} with narrowing of bounds.
Predictably, three intra-object bugs and one read-only bug could not be detected by the no-narrowing and only-writes versions of \mpx{} respectively.
ICC-\mpxshort{} detected only three bugs in total: in other cases programs failed due to \mpxshort-related issues (see \secref{compiler} and \secref{application}).
An interesting correlation emerged: the programs that contain real bugs are also the ones that break most often under \mpx.

As expected, AddressSanitizer found only three of these bugs---it checks bounds at the level of whole objects and cannot detect intra-object overflows.
SAFECode found bugs (2) and (3), the others either could not be detected due to coarse-grained granularity of bounds-checking or SAFECode could not compile the programs.
Unfortunately, SoftBound left bug (2) undetected and broke on other three programs with bugs: \code{ferret} and \code{x264} are multithreaded and thus not supported by SoftBound, and \code{perlbench} would not run correctly.

\subsection{\href{https://intel-mpx.github.io/usability/}{Usability}}
\label{sec:usability}

\begin{figure}[t]
	\centering
	\href{https://intel-mpx.github.io/usability/}{\includegraphics[scale=0.72]{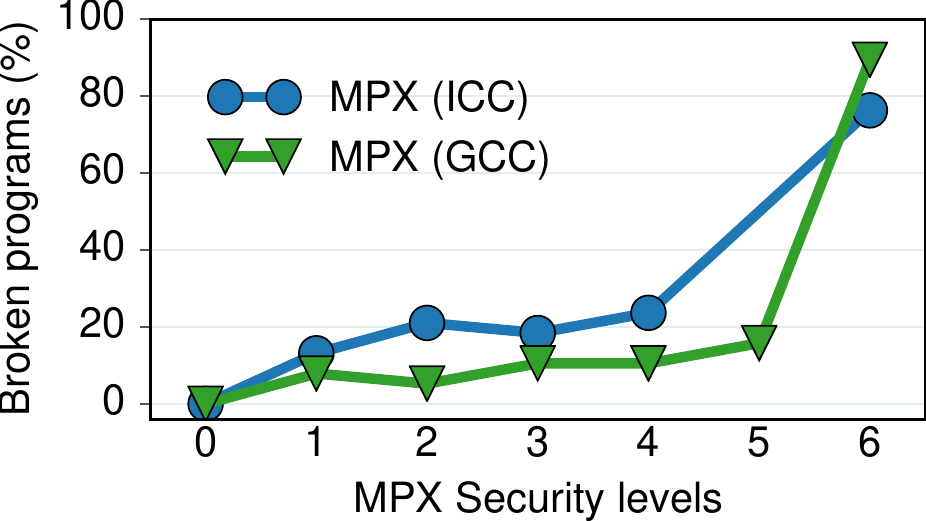}}
	\caption{Number of \mpxshort-broken programs rises with stricter \mpx{} protection rules (higher security levels). Level 4 is default.}
	\label{fig:usability}
\end{figure}

% tech. report version
As we showed in \secref{application}, some programs break under \mpx{} because they use unsupported C idioms or outright violate the C standard.
Moreover, as shown in \secref{compiler}, other programs even fail to compile or run due to internal bugs in the compiler \mpxshort{} passes (one case for GCC and 8 for ICC).

\figref{usability} highlights the \emph{usability} of \mpx, i.e., the number of \mpxshort-protected programs that fail to compile correctly and/or need significant code modifications.
Note that many programs can be easily fixed (see \tabref{application}); we do not count them as broken.
\mpxshort{} \emph{security levels} are based on our own classification and correspond to the stricter protection rules, where level 0 means unprotected native version and 6---the most secure \mpxshort{} configuration (see \secref{lessons}).
In total, our evaluation covers 38 programs from the Phoenix, PARSEC, and SPEC benchmark suites.

As can be seen, around $10\%$ of programs break already at the weakest level 1 of \mpx{} protection (without narrowing of bounds and protecting only writes).
At the highest security level 6 (with enabled \code{BNDPRESERVE}), most of the programs fail.

As for other approaches, \emph{no} programs broke under AddressSanitizer.
For SAFECode, around 70\% programs executed correctly (all Phoenix, half of PARSEC, and 3/4 of SPEC).
SoftBound---being a prototype implementation--showed poor results, with only simple programs surviving (all Phoenix, one PARSEC, and 6 SPEC).
These results roughly correspond to the ones in the original papers \cite{softbound09,safecode06}.

\begin{figure*}[t]
    \centering
    \href{https://intel-mpx.github.io/usability/\#usabilitytable}{\includegraphics[scale=0.9]{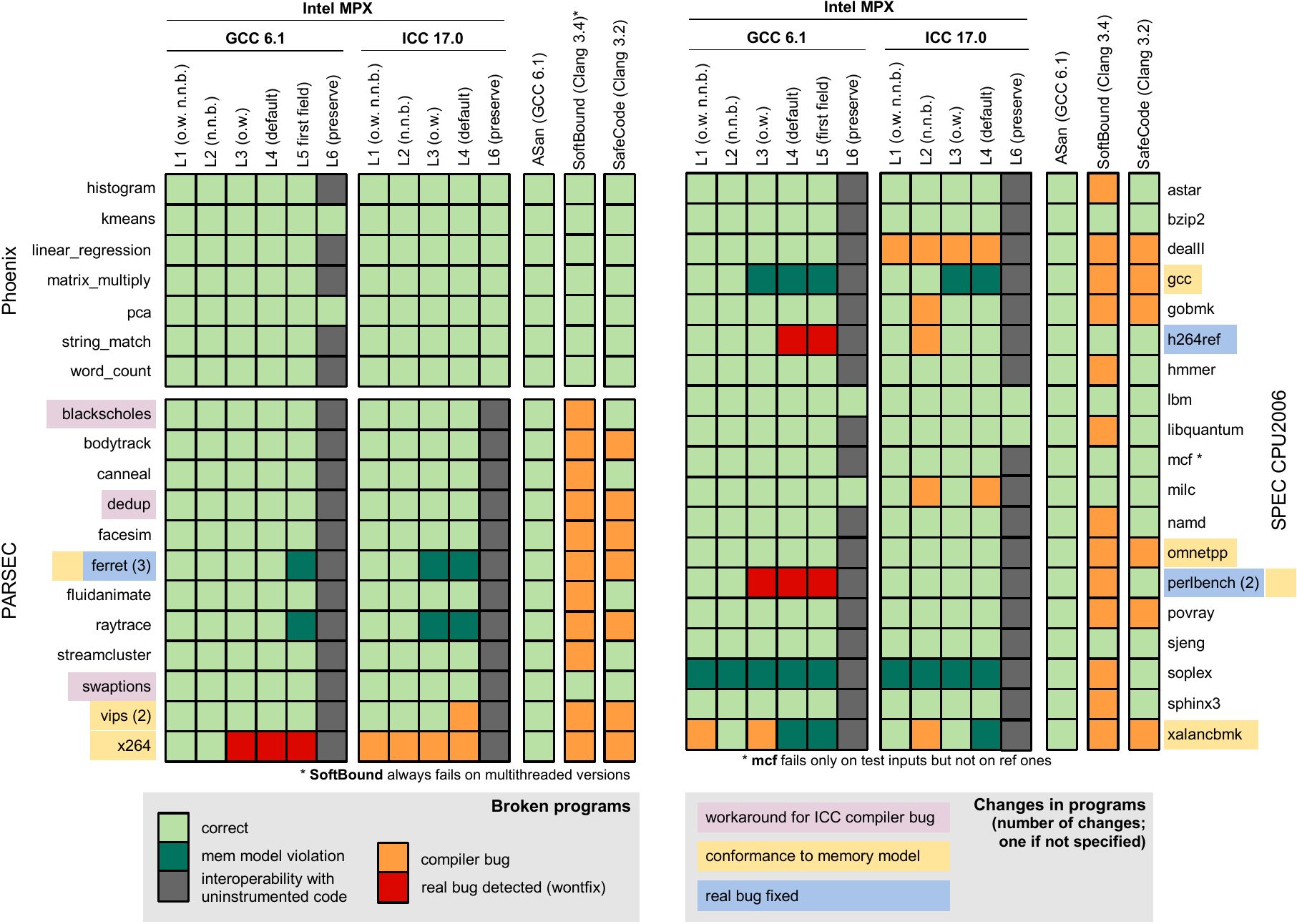}}
    \caption{All changes made to the programs under test as well as reasons why some programs break at compile- or run-time.}
    \label{fig:results_table}
\end{figure*}

\myparagraph{Encountered issues}
\figref{results_table} presents an overview of the issues we encountered during our experiments.

AddressSanitizer has no usability issues---by design it makes no assumptions on the C standard with respect to the memory model.
Also, it is the most stable tested product, fixed and updated with each new version of GCC and Clang.

On the contrary, SoftBound and SAFECode are research prototypes.
They work perfectly with very simple programs from Phoenix, but are not able to compile/run correctly the more complicated benchmarks from PARSEC and SPEC.
Moreover, SoftBound does not support multithreading, and any multithreaded program immediately fails under it.

Both GCC-\mpxshort{} and ICC-\mpxshort{} break most programs on Level 6 (with \code{BNDPRESERVE=1}).
This is because \code{BNDPRESERVE} does not clear bounds on pointers transferred from/to unprotected legacy libraries.
This means that any pointer returned from or modified by any legacy library (including C standard library) will almost certainly contain wrong bounds.
Because of this, 89\% of GCC-\mpxshort{} and 76\% of ICC-\mpxshort{} programs break.
These cases are represented as gray boxes.

Note that for Phoenix, GCC-\mpxshort{} fails in most cases while ICC-\mpxshort{} works correctly.
This is because of a slight difference in libc wrappers: all the failing programs use mmap64 function which is correctly wrapped by ICC-\mpxshort{} but ignored by GCC-\mpxshort{}.
Thus, in the GCC case, the newly allocated pointer contains no bounds which (under \code{BNDPRESERVE=1}) is treated as an out-of-bounds violation.

One can wonder why some programs still work even if interoperability with C standard library is broken.
The reason is that programs like kmeans, pca, and lbm require literally no external functions except malloc, memset, free, etc.---which are provided by the wrapper \mpxshort{} libraries.

Some programs break due to \emph{memory model violation}:
\begin{itemize}
    \item ferret and raytrace both have structs with the first field used to access other fields in the struct (a common practice that is actually disallowed by the C standard).
    ICC-\mpxshort{} disallows this behavior when bounds narrowing is enabled.
    GCC-\mpxshort{} allows such behavior by default and has a special switch to tighten it (\code{-fno-chkp-first-field-has-own-bounds}) which we classify as Level 5.
    \item gcc has its own complex memory model with bit-twiddling, type-casting, and other practices deprecated by the C standard.
    % This is why both GCC-\mpxshort{} and ICC-\mpxshort{} break as soon as bounds narrowing is enabled.
    \item soplex manually modifies pointers-to-object from one address to another using pointer arithmetic, without any respect towards pointer bounds.
    By design, \mpx{} cannot circumvent this violation of the C standard.
    (The same happens in mcf but only in one corner-case on test input.)
    \item xalancbmk performs a container-style subtraction from the base of a struct.
    This leads to GCC-\mpxshort{} and ICC-\mpxshort{} breaking when bounds narrowing is enabled.
    \item We also manually fixed some memory-model violations, e.g., flexible arrays with size 1 (\code{arr[1]}).
    These fixes are represented as yellow background.
\end{itemize}

In some cases, real bugs were detected (see also \secref{security}):

\begin{itemize}
    \item Three bugs in ferret, h264ref, and perlbench were detected and fixed by us.
    These fixes are represented as blue background.
    \item Three bugs in x264, h264ref, and perlbench were detected only by GCC-\mpxshort{} versions.
    These bugs are represented as red boxes.
    Note that ICC-\mpxshort{} missed bugs in h264ref and perlbench.
    Upon debugging, we noticed that ICC-\mpxshort{} narrowed bounds less strictly than GCC-\mpxshort{} and thus missed the bugs.
    We were not able to hunt out the root cause, but presume it is due to different memory layouts generated by GCC and ICC compilers.
\end{itemize}

In rare cases, we hit compiler bugs in GCC and ICC:

\begin{itemize}
    \item GCC-\mpxshort{} had only one bug, an obscure ``fatal internal GCC compiler error'' on only-write versions of xalancbmk.
    \item ICC-\mpxshort{} has an autovectorization bug triggered on some versions of vips, gobmk, h264ref, and milc.
    \item ICC-\mpxshort{} has a ``wrong-bounds through indirect call'' bug triggered on some versions of x264 and xalancbmk.
    \item ICC-\mpxshort{} has a bug we could not identify triggered on dealII.
    \item We also manually fixed all manifestations of the C99 VLA bug in ICC-\mpxshort{}.
    These bugs are represented as pink background.
\end{itemize}

\section{\href{https://intel-mpx.github.io/case-studies/}{Case Studies}}
\label{sec:casestudies}

To understand how \mpx{} affects complex real-world applications, we experimented with three case studies: Apache and Nginx web servers and Memcached memory caching system.
Similar to the previous section, we evaluated these programs along three dimensions: performance and memory overheads, security guarantees, and usability.

We compare default \mpx{} implementations of both GCC and ICC against the native version, as well as AddressSanitizer.
We were not able to compile any of the case studies under SoftBound and SAFECode: in most cases, the Configure scripts complained about an ``unsupported compiler'', and in one case (Apache under SoftBound) the compilation crashed due to an internal compiler error.
The native version we chose to show is GCC: native ICC and Clang versions have almost identical results, with an exception of Nginx explained later.
For the same reasons, we show only the GCC implementation of AddressSanitizer.

All experiments were performed on the same machines as in the previous section (\secref{study}).
One machine served as a server and a second one as clients, connected with a 1GB Ethernet cable and an actual bandwidth of $938$ Mbits/sec.
We configured all case studies to utilize all 8 cores of the server (details below).
For other configuration parameters, we kept their default values.
\remove{}

\begin{figure*}[t]
    \centering
    \href{https://intel-mpx.github.io/case-studies/}{\includegraphics[scale=0.75]{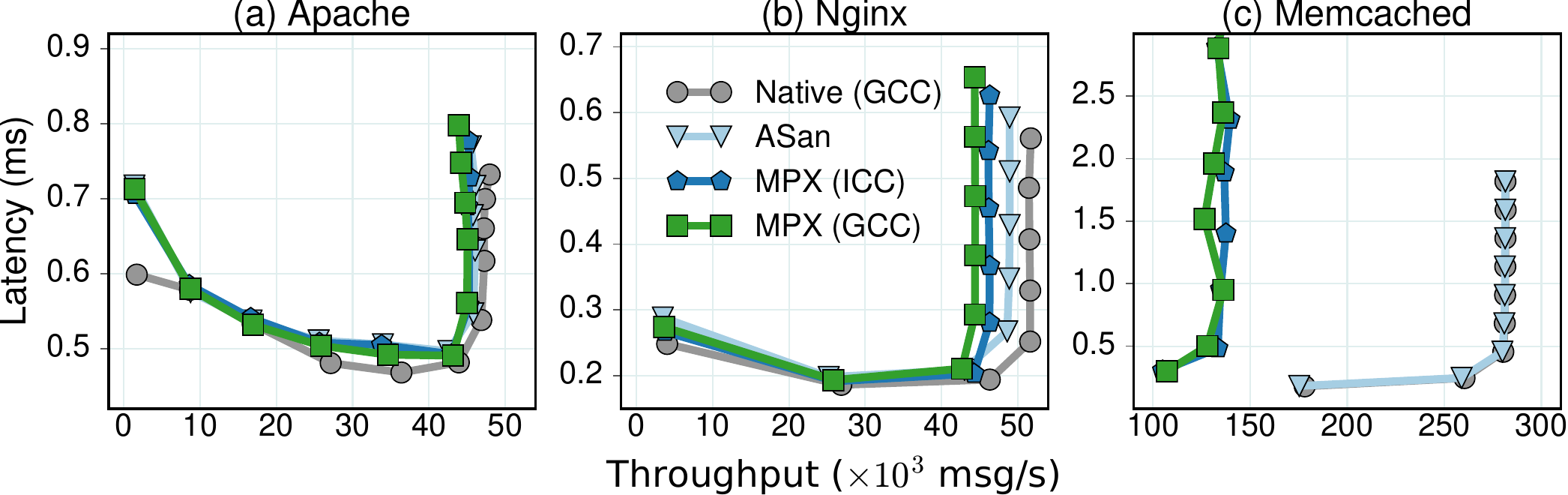}}
    \caption{Throughput-latency for (a) Apache web server, (b) Nginx web server, and (c) Memcached caching system.}
    \label{fig:casestudies}
\end{figure*}

All three programs were linked against their dependent libraries statically.
We opted for static linking to investigate the complete overhead of all components constituting each program.
\remove{}

\subsection{\href{https://intel-mpx.github.io/case-studies/\#apache}{Apache Web Server}}
\label{sec:apache}

For evaluation, we used Apache version 2.4.18 linked against OpenSSL 1.0.1f \cite{apache2016}.
This OpenSSL version is vulnerable to the infamous Heartbleed bug which allows the attacker to leak confidential information such as secret keys and user passwords in plain-text \cite{heartbleed2016}.
Since both AddressSanitizer and \mpx{} do not support inline assembly, we disabled it for all builds of Apache.
To fully utilize the server, we used the default configuration of Apache's MPM event model.

The classic ab benchmark was run on a client machine to generate workload, constantly fetching a static 2.3K web-page via HTTP, with a KeepAlive feature enabled.
To adapt the load, we increased the number of simultaneous requests at a time.

Unfortunately, while testing against Heartbleed, we discovered that ICC-\mpxshort{} suffers from a run-time Intel compiler bug\footnote{\url{https://software.intel.com/en-us/forums/intel-c-compiler/topic/700550}} in the \code{x509\_cb} OpenSSL function, leading to a crash of Apache.
This bug triggered only on HTTPS connections, thus allowing us to still run performance experiments on ICC-\mpxshort{}.

\myparagraph{Performance}
As \figref{casestudies}a shows, GCC-\mpxshort{}, ICC-\mpxshort{}, and AddressSanitizer all show minimal overheads, achieving $95.3\%$, $95.7\%$, and $97.5\%$ of native throughput.
Overhead in latency did not exceed $5\%$.
Such good performance is explained by the fact that our experiment was limited by the network and not CPU or memory.
(We observed around $480-520\%$ CPU utilization in all cases.)

In terms of memory usage (\tabref{casestudies_memory}), AddressSanitizer exhibits an expected $3.5\times$ overhead.
In contrast, \mpx{} variants have dramatic $12.8\times$ increase in memory consumption.
This is explained by the fact that Apache allocates an additional $1$MB of pointer-heavy data per each client, which in turn leads to the allocation of many Bounds Tables.

% -*- root: ../main.tex -*-
\begin{table}[t]
	\small
	\centering
	\begin{tabular}{p{2.2cm} r r r}
		& \bfseries Apache & \bfseries Nginx & \bfseries Memcached \\
		\hline
		\hline\\[-8pt] 
		Native     & 9.4           & 4.3           & 73 \\
		MPX        & \bfseries 120 & 18            & \bfseries 352 \\
		ASan       & 33            & \bfseries 380 & 95 \\
		\hline
	\end{tabular}
	\caption{Memory usage (MB) for peak throughput. (GCC-MPX and ICC-MPX showed identical results.)}
	\label{tab:casestudies_memory}
\end{table}

\myparagraph{Security}
For security evaluation, we exploited the infamous Heartbleed bug \cite{heartbleed2016,heartbleeddetails2014}.
In a nutshell, Heartbleed is triggered when a maliciously crafted TLS heartbeat message is received by the server.
The server does not sanity-check the length-of-payload parameter in the message header, thus allowing \code{memcpy} to copy the process memory's contents in the reply message.
In this way, the attacker can read confidential memory contents.

AddressSanitizer and GCC-\mpxshort{} detect Heartbleed\footnote{The actual situation with Heartbleed is more contrived. OpenSSL uses its own memory manager which partially bypasses the wrappers around \code{malloc} and \code{mmap}. Thus, in reality memory-safety approaches find Heartbleed only if the length parameter is greater than 32KB (the granularity at which OpenSSL allocates chunks of memory for its internal allocator) \cite{heartbleedmemory2014}.}.

\subsection{\href{https://intel-mpx.github.io/case-studies/\#nginx}{Nginx Web Server}}
\label{sec:nginx}

We tested Nginx version 1.4.0---the last version with a stack buffer overflow vulnerability \cite{nginx2016}.
Nginx was configured with the ``autodetected'' number of worker processes to load all cores and was benchmarked against the same ab benchmark as Apache.
ab was also used as a client.

To successfully run Nginx under GCC-\mpxshort{} with narrowing of bounds, we had to manually fix a variable-length array \code{name[1]} in the \code{ngx\_hash\_elt\_t} struct to \code{name[0]}.
However, ICC-\mpxshort{} with narrowing of bounds still refused to run correctly, crashing with a false positive in \code{ngx\_http\_merge\_locations} function.
In a nutshell, the reason for this bug was a cast from a smaller type, which rendered the bounds too narrow for the new, larger type.
Note that GCC-\mpxshort{} did \emph{not} experience the same problem because it enforces the first struct's field to inherit the bounds of the whole object by default---in contrast to ICC-\mpxshort{} which takes a more rigorous stance.
For the following evaluation, we used the version of ICC-\mpxshort{} with narrowing of bounds disabled.

\myparagraph{Performance}
With regards to performance (\figref{casestudies}b), Nginx has a similar behavior to Apache.
AddressSanitizer reaches $95\%$ of native throughput, while GCC-\mpxshort{} and ICC-\mpxshort{} lag behind with $86\%$ and $89.5\%$ respectively.
Similar to Apache, this experiment was network-bound, with CPU usage of $225\%$ for native, $265\%$ for \mpx, and $300\%$ for AddressSanitizer.
(CPU usage numbers prove that HW-assisted approaches impose less CPU overheads.)

As a side note, Nginx has predictable behavior only under GCC.
Native ICC version reaches only $85\%$ of the GCC's throughput, and native Clang only $90\%$.
Even more surprising, the ICC-\mpxshort{} version performed $5\%$ \emph{better} than native ICC; similarly, the AddressSanitizer-Clang version was $10\%$ \emph{better} than native Clang.
We are still investigating the reasons for this unexpected behavior.

As for memory consumption (\tabref{casestudies_memory}), the situation is opposite as with Apache: \mpx{} variants have a reasonable $4.2\times$ memory overhead, but AddressSanitizer eats up $88\times$ more memory (it also has $625\times$ more page faults and $13\%$ more LLC cache misses).
But then why \mpx{} is slower than AddressSanitizer if their memory characteristics indicate otherwise?
The reason for the horrifying AddressSanitizer numbers is its ``quarantine'' feature---AddressSanitizer employs a special memory management system which avoids re-allocating the same memory region for new objects, thus decreasing the probability of temporal bugs such as use-after-free.
Instead, AddressSanitizer marks the used memory as ``poisoned'' and requests new memory chunks from the OS (this explains huge number of page faults).
Since native Nginx recycles the same memory over and over again for the incoming requests, AddressSanitizer experiences huge memory blow-up.
When we disabled the quarantine feature, AddressSanitizer used only $24$MB of memory.

Note that this quarantine problem does not affect performance.
Firstly, Nginx is network-bound and has enough spare resources to hide this issue.
Secondly, the rather large overhead of allocating new memory hides the overhead of requesting new chunks from the OS.

\myparagraph{Security}
To evaluate security, the bug under test was a stack buffer overflow CVE-2013-2028 that can be used to launch a ROP attack \cite{nginxbug2013}.
Here, a maliciously crafted HTTP request forces Nginx to erroneously recognize a signed integer as unsigned.
Later, a \code{recv} function is called with the overflown size argument and the bug is triggered.

Perhaps surprisingly, AddressSanitizer detects this bug, but both versions of \mpx{} \emph{do not}.
The root cause is the run-time wrapper library: AddressSanitizer wraps \emph{all} C library functions including \code{recv}, and the wrapper---not the Nginx instrumented code---detects the overflow.
In case of both GCC-\mpxshort{} and ICC-\mpxshort{}, only the most widely used functions, such as \code{memcpy} and \code{strlen}, are wrapped and bounds-checked.
That is why when \code{recv} is called, the overflow happens in the unprotected C library function and goes undetected by \mpx.

This highlights the importance of full protection---not only protecting the program's own code, but also writing wrappers around all unprotected libraries used by the program.
Another interesting aspect is that this overflow bug is read-only and cannot be caught by write-only protection.
No matter how tempting it may sound to protect only writes, one must remember that buffer-overread vulnerabilities will slip away from such low-overhead bounds checking.

\subsection{\href{https://intel-mpx.github.io/case-studies/\#memcached}{Memcached Caching System}}
\label{sec:memcached}

Lastly, we experimented with Memcached version 1.4.15 \cite{memcached2004}.
This is the last version susceptible to a simple DDoS attack \cite{memcachedbug2011}.
In all experiments, Memcached was run with 8 threads to fully utilize the server.
For the client we used a memaslap benchmark from libmemcached with a default configuration ($90\%$ reads of average size $1700$B, $10\%$ writes of average size $400$B).
We increased the load by adapting the concurrency number.

After some vexing debugging experiences with Nginx and Apache, we were pleased to experience no issues instrumenting Memcached with GCC-\mpxshort{} and ICC-\mpxshort{}.

% -*- root: ../main.tex -*-
\begin{table*}

\footnotesize
\centering
\setlength\tabcolsep{3pt} % default value: 6pt

\begin{tabular}{@{\extracolsep{3pt}}p{0.5cm} p{5.3cm} p{2.8cm} >{\centering\arraybackslash}p{0.6cm} >{\centering\arraybackslash}p{0.6cm} >{\centering\arraybackslash}p{0.6cm} >{\centering\arraybackslash}p{0.6cm} >{\centering\arraybackslash}p{0.5cm} >{\centering\arraybackslash}p{0.4cm} >{\centering\arraybackslash}p{0.5cm} >{\centering\arraybackslash}p{0.5cm}}
%[-8pt]
%\hline
\noalign{\smallskip}
& & & \multicolumn{2}{c}{\bfseries RIPE attacks} & \multicolumn{2}{c}{\bfseries Unfound bugs} & \multicolumn{2}{c}{\bfseries Broken} & \multicolumn{2}{c}{\bfseries Perf ($\times$)}\\
\cline{4-5}\cline{6-7}\cline{8-9}\cline{10-11}
\noalign{\vskip 1pt}
\bfseries Level & \bfseries Description & \bfseries Detects & \bfseries GCC & \bfseries ICC & \bfseries GCC & \bfseries ICC & \bfseries GCC & \bfseries ICC & \bfseries GCC & \bfseries ICC \\
 \noalign{\vskip -9pt}
\hline
\hline
\noalign{\vskip 1pt}
\bfseries 0 & native program (no protection)                              & \hspace*{0.1cm} ---                      & 64 & 34 & 6 & 3 & 0  & 0 & 1.00 & 1.00 \\
\bfseries 1 & \mpxshort{} only-writes and no narrowing of bounds               & inter-object overwrites                  & 14 & 14 & 3 & 0 & 3  & 5 & 1.29 & 1.18  \\
\bfseries 2 & \mpxshort{} no narrowing of bounds                               & \hspace*{0.1cm} + inter-object overreads & 14 & 14 & 3 & 0 & 2  & 8 & 2.39 & 1.46 \\
\bfseries 3 & \mpxshort{} only-writes and narrowing of bounds                  & all overwrites*                          & 14 & 0  & 2 & 0 & 4  & 7 & 1.30 & 1.19 \\
\bfseries 4 & \mpxshort{} narrowing of bounds (default)                        & \hspace*{0.1cm} + all overreads*         & 14 & 0  & 0 & 0 & 4  & 9 & 2.52 & 1.47 \\
\bfseries 5 & \hspace*{0.1cm} + \code{fchkp-first-field-has-own-bounds}*  & \hspace*{0.1cm} + all overreads          & 0  & -- & 0 & --& 6  & --& 2.52 & -- \\
\bfseries 6 & \hspace*{0.1cm} + \code{BNDPRESERVE=1} (protect all code)   & all overflows                            & 0  & 0  & 0 & 0 & 34 & 29& --   & -- \\
\\ [-7pt]
            & AddressSanitizer \cite{asan12}                              & inter-object overflows                   & 12 & -- & 3 & --& 0  & --& 1.55  & --\\
%\\ [-18pt]
\hline
\\ [-8pt]
\multicolumn{11}{r}{\scriptsize{* except intra-object overflows through the first field of struct, level 5 removes this limitation (only relevant for GCC version)}}
\\
\end{tabular}

\caption{The summary table with our classification of \mpx{} security levels---from lowest L1 to highest L6---highlights the trade-off between \underline{security} (number of unprevented 	\emph{RIPE attacks} and other \emph{Unfound bugs} in benchmarks), \underline{usability} (number of \mpxshort-\emph{Broken} programs), and \underline{performance overhead} (average \emph{Perf} overhead w.r.t. native executions). AddressSanitizer is shown for comparison in the last row.}
\label{tab:summary}
\vspace{-2mm}
\end{table*}

\myparagraph{Performance}
Performance-wise, Memcached turned out to be the worst case for \mpx{} (see \figref{casestudies}c).
While AddressSanitizer performs on par with the native version, both GCC-\mpxshort{} and ICC-\mpxshort{} achieved only $48-50\%$ of maximum native throughput.

In case of native and AddressSanitizer, performance of Memcached was limited by network.
But it was not the case for \mpx: Memcached exercised only $70\%$ of the network bandwidth.
The memory usage numbers in \tabref{casestudies_memory} help understand the bottleneck of \mpx.
While AddressSanitizer imposed only $30\%$ memory overhead, both \mpx{} variants used $350$MB of memory ($4.8\times$ more than native).
This huge memory overhead broke cache locality and resulted in $5.4\times$ more page faults and $10-15\%$ LLC misses, making \mpx{} versions essentially memory-bound.
(Indeed, the CPU utilization never exceeded $320\%$.)

\myparagraph{Security}
For security evaluation, we used a CVE-2011-4971 vulnerability \cite{memcachedbug2011}.
In this denial-of-service attack, a specially crafted packet is received by the server and passed to the handler (\code{conn\_nread}) which tries to copy all packet's contents into another buffer via the \code{memmove} function.
However, due to the integer signedness error in the size argument, \code{memmove} tries to copy gigabytes of data and quickly segfaults.
All approaches---AddressSanitizer, GCC-\mpxshort{}, and ICC-\mpxshort{}---detected buffer overflow in the affected function's arguments and stopped the execution.

% -*- root: main.tex -*-
%!TEX root = main.tex
\section{Lessons Learned}
\label{sec:lessons}

\tabref{summary} summarizes the results of our work.
For convenience, we introduce six \emph{\mpx{} security levels} to highlight the trade-offs between security, usability, and performance.

In general, \mpx{} is a promising technology: it provides the strongest possible security guarantees against spatial errors, it instruments most programs transparently and correctly, its ICC incarnation has moderate overheads of \textasciitilde50\%, it can interoperate with unprotected legacy libraries, and its protection level is easily configurable.
However, our evaluation indicates that it is not yet ready for widespread use because of the following issues:

\myparagraph{Lesson 1: New instructions are not as fast as expected}
First, current Skylake processors perform bounds checking mostly sequentially.
Our microbenchmarks indicate this is caused by contention of bounds-checking instructions on one execution port.\footnote{We project that, if this functionality would be available on more ports, \mpx{} would be able to use instruction parallelism to a higher extent and the overheads would be lower.}
Secondly, loading/storing bounds registers from/to memory involves costly two-level address translation, which can contribute a significant share to the overhead.
Together, these two issues lead to tangible runtime overheads of \textasciitilde50\% even with all optimizations applied (in the ICC case).
%We, however, do not know any viable solution to this problem.

\myparagraph{Lesson 2: The supporting infrastructure is not mature enough}
\mpx{} support is available for GCC and ICC compilers.
At the compiler level, GCC-\mpxshort{} has severe performance issues (\textasciitilde$150\%$) whereas ICC-\mpxshort{} has a number of compiler bugs (such that $10\%$ of programs broke in our evaluation).
At the runtime-support level, both GCC and ICC provide only a small subset of function wrappers for the C standard library, thus not detecting bugs in many libc functions.

\myparagraph{Lesson 3: \mpx{} provides no temporal protection}
Currently, \mpx{} protects only against spatial (out-of-bounds accesses) but not temporal (dangling pointers) errors.
All other tested approaches---AddressSanitizer, SoftBound, and SAFECode---guarantee some form of temporal safety.
We believe \mpx{} can be enhanced for temporal safety without harming performance, similar to SoftBound.

\myparagraph{Lesson 4: \mpx{} does not support multithreading}
An \mpxshort-protected multithreaded program can have both false positives (false alarms) and false negatives (missed bugs and undetected attacks).
Until this issue is fixed---either at the software or at the hardware level---\mpx{} cannot be considered safe in multithreaded environments.
Unfortunately, we do not see a simple fix to this problem that would not affect performance adversely.

\myparagraph{Lesson 5: \mpx{} is not compatible with some C idioms}
\mpx{} imposes restrictions on allowed memory layout which conflict with several widespread C programming practices, such as intra-structure memory accesses and custom implementation of memory management.
This can result in unexpected program crashes and is hard to fix; we were not able to run correctly 8--13\% programs (this would require substantial code changes).

%\smallskip
\textbf{In conclusion}, we believe that \mpx{} has a potential for becoming the memory protection tool of choice, but currently, AddressSanitizer is the only production-ready option.
Even though it provides weaker security guarantees than the other techniques, its current implementation is better in terms of performance and usability.
SoftBound and SAFECode are research prototypes and they have issues that restrict their usage in real-world applications (although SoftBound provides higher level of security).

%Both implementations of \mpx{} do not support C programming idioms to the full extent, which causes a significant number of false positives in complex programs.
%GCC implementation is less susceptible to them, but it comes at a cost of worse performance.
We expect that most identified issues with \mpx{} will be fixed in future versions.
Still, support for multithreading and restrictions on memory layout are inherent design limitations of \mpx{} which would require sophisticated solutions, which would in turn negatively affect performance.
We hope our work will help practitioners to better understand the benefits and caveats of \mpx{}, and researchers---to concentrate their efforts on those issues still waiting to be solved.

All sources of our experiments can be found in the \href{https://github.com/OleksiiOleksenko/intel_mpx_explained}{public repository}.

% -*- root: main.tex -*-
%!TEX root = main.tex
\section{Acknowledgments}
\label{sec:acks}

We would like to thank the developer of the GCC-\mpxshort{} pass Ilya Enkovich, the authors of AddressSanitizer (Konstantin Serebryany and Alexander Potapenko), SoftBound (Santosh Nagarakatte and Milo Martin), and SAFECode (John Criswell) for the provided help with their tools and for answering our questions.
We also thank the anonymous reviewers, Bohdan Trach, Sergei Arnautov, and Franz Gregor for their insightful reviews, helpful comments and proof-reading.

%\footnotesize
{\footnotesize \bibliographystyle{acm}
\bibliography{refs.bib}}

\end{document}